\documentclass[twocolumn]{aastex63}

\usepackage{color}
\usepackage{mathrsfs}
\usepackage{graphicx}
\usepackage{cases}
\usepackage{txfonts}
\usepackage{epsfig}
\usepackage{multirow}
\usepackage{bm}
\usepackage{aas_macro}

\def\be{\begin{equation}}
\def\ee{\end{equation}}

\def\bea{\begin{eqarray}}
\def\eea{\end{eqarray}}
\def\bs{\bm}

\begin{document}

\title{Scalar quadratic maximum likelihood estimators for the CMB cross power spectrum}

\author{Jiming Chen}\email{chenjm94@mail.ustc.edu.cn}
\affiliation{CAS Key Laboratory for Researches in Galaxies and Cosmology, Department of Astronomy, University of Science and Technology of China, Chinese Academy of Sciences, Hefei, Anhui 230026, China}
\affiliation{School of Astronomy and Space Science, University of Science and Technology of China, Hefei, 230026, China}

\author{Shamik Ghosh}\email{shamik@ustc.edu.cn}
\affiliation{CAS Key Laboratory for Researches in Galaxies and Cosmology, Department of Astronomy, University of Science and Technology of China, Chinese Academy of Sciences, Hefei, Anhui 230026, China}
\affiliation{School of Astronomy and Space Science, University of Science and Technology of China, Hefei, 230026, China}

\author{Wen Zhao}\email{wzhao7@ustc.edu.cn}
\affiliation{CAS Key Laboratory for Researches in Galaxies and Cosmology, Department of Astronomy, University of Science and Technology of China, Chinese Academy of Sciences, Hefei, Anhui 230026, China}
\affiliation{School of Astronomy and Space Science, University of Science and Technology of China, Hefei, 230026, China}


\shorttitle{Fast scalar QML estimators for the CMB $B$-mode power spectrum}
\shortauthors{Chen, J., Ghosh, S., Zhao, W.}

\begin{abstract}
Estimating the cross-correlation power spectra of cosmic microwave background (CMB), in particular, the $TB$ and $EB$ spectra, is important for testing parity symmetry in cosmology and diagnosing insidious instruments {systematics}.  The Quadratic Maximum Likelihood (QML) estimator provides the optimal estimates of power spectra, but it is computationally very expensive. The hybrid pseudo-$C_\ell$ estimator is computationally fast but performs poorly on large scales. As a natural extension of previous work \citep{chen2021fast}, in this article, we present a new unbiased estimator based on the Smith-Zaldarriaga (SZ) approach of $E$-$B$ separation and scalar QML approach to reconstruct the cross-correlation power spectrum, called QML-SZ {estimator.} Our new estimator relies on the ability to construct scalar maps, {which allows us to use a scalar QML estimator to obtain the cross-correlation} power spectrum. By reducing the pixel number and algorithm complexity, the computational cost is nearly one order of magnitude smaller and the running time is nearly two orders of magnitude faster in the test situations. 
\end{abstract}

\keywords{Cosmic microwave background, Polarization, Power spectrum, Quadratic maximum likelihood}

\section{Introduction}
\label{intro}
In the past two decades, a series of the cosmic microwave background (CMB) experiments, e.g. DASI \citep{DASI}, WMAP \citep{WMAP1, WMAP2, WMAP3}, BOOMERanG \citep{2006ApJ...647..813M}, QUAD \citep{QUAD}, BICEP \citep{BICEP}, QUIET \citep{2012ApJ...760..145Q}, ACT \citep{2014JCAP...10..007N}, Planck \citep{2014A&A...571A..16P}, SPTpol \citep{2018ApJ...852...97H}, have already provided large amounts of high quality data, responsible for obtaining tight constraints on the cosmological parameters. As the study of cosmology into the age of precision, the research of CMB cross power spectra becomes possible and plays a more important role, both for the data characterization and their scientific interpretation.

In the early cosmological scenario, quantum fluctuations produce primordial density and primordial gravitational waves.  Both contribute to the CMB temperature anisotropy, the latter also produces a distinguishable feature of CMB polarization. According to the standard cosmological model, primordial gravitational waves can not only generate auto-correlation $TT$, $EE$ and $BB$ power spectrum of CMB, but also $TE$ power spectrum. The classic cosmological model believes that the physical mechanism in the process of photon propagation is parity invariant, which results in the spectra of $EB$ and $TB$ vanish \citep{zaldarriaga-b-mode, kamionkowski97,krauss2010primordial,garcia2011primordial}. The information of these power spectrums can be used to probe the primordial fluctuations. $TB$ and $EB$ power spectrum are 
good null test and can be used to detect the presence of an instrument and/or astrophysical system effects \citep{2003PhRvD..67d3004H, 2010PhRvD..81f3512Y}. In addition, some non-standard cosmological mechanisms that could produce non-vanishing cross-spectra and reconstruct the cross power spectrum is also of great significance in checking some parity-violating interactions and go beyond the standard models \citep{1999PhRvL..83.1506L, PhysRevLett.96.221302,PhysRevD.78.103516,
PhysRevD.87.103512,
zhao2014a, zhao2014b,PhysRevD.88.063508, PhysRevD.101.043528}. And these tests may have far-reaching consequences for our understanding of the Universe. 

Associated with CMB experimental developments, many investigations of techniques have been developed to reconstruct cross-correlation power spectrum from maps of the partial CMB sky. A standard approach presented in \citep{2002ApJ...567....2H,2005MNRAS.358..833T} is the most straightforward way to construct the power spectrum for partial sky situation, which usually be used to estimate temperature multipoles and $E$-mode multipoles. {Since in the standard approach there is} no explicit correction for the $E$-to-$B$ leakage, the performance of using this approach to estimate $B$-mode multipoles is poor. To solve the problem that exists in the standard approach, several extensions of the standard pseudo-$C_{\ell}$ (PCL) methods \citep{ 2003MNRAS.343..559H, ebmixture9, 2007PhRvD..76d3001S, 2010PhRvD..82b3001Z, 2010A&A...519A.104K, 2011A&A...531A..32K, 2012PhRvD..86g6005G, 2019PhRvD.100b3538L,Ghosh2020} have been proposed. These methods reconstruct the power spectra by inverting the linear system relating the full sky power to the power from the incomplete sky. These methods 
based on the fast spherical harmonic transforms with the advantage of speeding up their computation. The method proposed in \citep{ebmixture9, 2007PhRvD..76d3001S} (hereafter the SZ method) was shown to be the PCL estimator \citep{ferte2013} with the smallest errors. In our paper, we adopt a hybrid approach \citep{2012PhRvD..86g6005G} to reconstruct $TB$ and $EB$, where the $T$-mode and $E$-mode multipoles are obtained from the standard estimator and the $B$-mode multipoles are obtained from the SZ estimator.


 The Quadratic Maximum Likelihood (QML) method \citep{2001PhRvD..64f3001T}, which is a pixel-based estimator, provides another way to solve the $E$-$B$ mixing problem. It has the advantage of minimizing spectra uncertainties. However, it involves matrix inversions and multiplications which significantly increases the calculation time and the demand for computational memory. 
 
 Combining the advantages of the above two methods, we propose another new method to reconstruct the large-scale cross-correlation power spectrum: the QML-SZ method. The QML-SZ method uses the SZ-method to derive the pure $E$-mode map $\mathcal{E}(\hat{n})$ and the pure $B$-mode map $\mathcal{B}(\hat{n})$ from Stokes $Q$ and $U$ maps, which can be ultimately treated as scalar fields, as for $T$-mode map ${T}(\hat{n})$ it has been a scalar map. These scalar maps enabling us to use the QML method developed for CMB temperature maps to calculate $TE$, $TB$ and $EB$ power spectrum. Since we adopt the scalar mode QML method to reconstruct the cross-correlation power spectrum, the number of pixels drops to $1/3$ of the standard QML method. This means that the computational running time will \textbf{be} greatly shorter than the standard QML estimator and drastically reduce our computational requirements.

This paper is organized as follows. In Section~\ref{sec:notation_and_conventions}, we present our conventions and notation for describing CMB fields on the sphere. In Section~\ref{sec:power_spectrum_estimators}, we review the three pseudo-$C_\ell$ estimators and SZ estimator first, and then introduce the scalar mode QML method and combine it with the hybrid PCL method to construct the QML-SZ estimator. The simulation setup and details can be found in Section ~\ref{sec:SIMULATION_SETUP}. In Section~\ref{sec:realistic_examples}, we apply these methods to realistic situations and make a comprehensive comparison of their performance. Conclusions and discussions are given in Section~\ref{sec:conclusion}. Appendix \ref{apdx:TE} show the results of $TE$ power spectrum.

\section{NOTATION AND CONVENTIONS}
\label{sec:notation_and_conventions}
In this section we will briefly summarise the notations and definitions used in this paper. The CMB temperature fluctuation on the sphere is a scalar field, with fluctuations $\Delta T(\hat{n})$ at the level of $10^{-5}$ of the average value $T = 2.725 K$. For full sky observations, the CMB temperature fluctuations can be expanded in spherical harmonics as,
 \begin{equation}
    \Delta T(\hat{n})=\sum_{\ell m} T_{ \ell m}Y_{\ell m}(\hat{n}),
 \end{equation}
 where $\hat{n}$ denotes the line-of sight, $Y_{\ell m}(\hat{n})$ are the spherical harmonics and $T_{ \ell m}$ are the corresponding coefficients. The linearly polarized CMB polarization field does not contain a circular polarization component. Therefore, it can be characterized by Stokes parameters, $Q$ and $U$. We can define ${}_{\pm}P(\hat{n})$ as follows
\begin{equation}
\label{PQU}
	{}_{\pm}P(\hat{n})=Q(\hat{n})\pm iU(\hat{n}).
\end{equation}
The definition of $Q$ and $U$ fields are coordinate dependent, so the polarization fields ${}_{\pm}P(\hat{n})$, behave as a spin-($\pm2$) field. These spin-($\pm2$) fields can be also expanded in spin-weighted spherical harmonics, ${}_{\pm s}Y_{\ell m}$ with spin $s=\pm2$ (detailed expressions can be found in \citet{Newman1966, ebmixture0}), as \citep{seljak1996}:
\begin{equation}
{}_{\pm}P(\hat{n})=\sum_{\ell m} a_{\pm2,\ell m}~{}_{\pm 2}Y_{\ell m}(\hat{n}).
\end{equation}
The polarization field can be decomposed into $E$- and $B$-mode parts, which can be expressed as linear combinations of $a_{\pm 2,\ell m}$.
\begin{eqnarray}
E_{ \ell m} &\equiv& -\frac{1}{2}[a_{2,\ell m}+a_{-2,\ell m}] \nonumber \\
B_{ \ell m}& \equiv& -\frac{1}{2i}[a_{2,\ell m}-a_{-2,\ell m}]. 
\label{eq_blm} 
\end{eqnarray}

One can construct the scalar $E$-mode, and pseudoscalar $B$-mode fields as,
 \begin{eqnarray}
 E(\hat{n}) \equiv \sum_{\ell m}E_{\ell m}Y_{\ell m}(\hat{n}), \qquad
 B(\hat{n}) \equiv \sum_{\ell m}B_{\ell m} Y_{\ell m}(\hat{n}).
 \end{eqnarray}
Finally, the power spectrum estimate can be obtained as follows \citep{grishchuk1997,zhao2009}
\begin{eqnarray}
\hat C^{XZ}_{\ell } \equiv \frac{1}{2\ell +1} \sum_m X_{\ell m}Z^{*}_{\ell m},
\end{eqnarray}
where $X$, $Z \in (T,E,B)$. 

\section{POWER SPECTRUM ESTIMATORS} 
\label{sec:power_spectrum_estimators}

\subsection{Pseudo$-C_{\ell}$ estimator}
\label{CMB $B$-mode decomposition}
In this subsection, we will introduce standard and pure harmonic coefficients for incomplete sky coverage. The $E$- and $B$-mode decomposition is not unique on an incomplete sky, which leads to leakage from $E$-to-$B$ and $B$-to-$E$.  The standard harmonic coefficient relations have no explicit correction for this leakage while the pure method corrects for the leakage problem. Since the $E$ modes have much larger power than $B$ modes, the pure method is only used in the $B$-mode case. So, the PCL estimator in this work uses standard simple harmonic calculation for $T$, $E$ modes, and pure method for $B$ modes. We will first summarize these definitions before defining the PCL estimator relations. 
\subsubsection{Standard harmonic coefficient definitions}
\label{standard pseudo multipoles}
For an incomplete sky observation, defined by a binary mask $W$, we can intuitively define the partial sky $T$-, $E$- and $B$-mode harmonic coefficients (indicated by overhead tilde) as:
\begin{eqnarray}
    \tilde T_{\ell m} &=& \int \Delta T\, W\, Y^*_{\ell m}d\hat{n}, \\
    \tilde E_{\ell m} &=& -\frac{1}{2}\int  W\, \left[{}_+P\, {}_{+2}Y^*_{\ell m} + {}_-P\, {}_{-2}Y^*_{\ell m}\right]d\hat{n}, \\
    \tilde B_{\ell m} &=& -\frac{1}{2i}\int W\, \left[{}_+P\, {}_{+2}Y^*_{\ell m} - {}_-P\, {}_{-2}Y^*_{\ell m}\right]d\hat{n}.
\end{eqnarray}
The relationship between these partial sky coefficients and the full sky coefficients can be expressed as,
\begin{eqnarray}
\tilde T_{\ell m} &=& \sum_{\ell' m'} K_{\ell m \ell' m'}^{TT}T_{\ell' m'}, \nonumber\\
\tilde E_{\ell m} &=& \sum_{\ell' m'}\left[ K_{\ell m \ell' m'}^{EE}E_{\ell' m'} + iK_{\ell m \ell' m'}^{EB}B_{\ell' m'}\right], \nonumber\\
\tilde B_{\ell m} &=& \sum_{\ell' m'}\left[- iK_{\ell m \ell' m'}^{BE}E_{\ell' m'} + K_{\ell m \ell' m'}^{BB}B_{\ell' m'} \right].
\label{std_coupling_correalation}
\end{eqnarray}
The coupling matrices, $K_{\ell m \ell' m'}^{XY}$, can derived from the definitions, and the full form of these matrices can be found in \citet{ferte2013}. 

\subsubsection{Pure harmonic coefficient definitions}
\label{pure pseudo multipoles}
In any CMB experiment we perform $E$-$B$ decomposition on an incomplete sky due to foreground masking or survey footprint. As we stated before this leads to leakage from one type of polarization mode to another. Since the $B$ modes are orders of magnitude smaller than the $E$-mode signal, the $E$-to-$B$ leakage is a critical problem for CMB polarization experiments. Various methods have been proposed in the literature to avoid $E$-$B$ leakage problem, such as \cite{ebmixture0, 2011PhRvD..83h3003B, 2003PhRvD..68h3509L, 2009ApJ...706.1545C, 2013MNRAS.435.2040L, 2009PhRvD..79l3515G,ebmixture9, 2007PhRvD..76d3001S, 2010PhRvD..82b3001Z, 2010A&A...519A.104K,larissa2016,larissa2017,Ghosh2020}. The pure method (also called the SZ method), proposed in \citet{ebmixture9} and \citet{2007PhRvD..76d3001S}, has been shown \citep{ferte2013} to have the best performance in reducing error bars. The key of this approach is to apply the spin-raising and spin-lowering operators \citep{Newman1966}, $\eth$ and $\bar \eth$, on ${}_\pm P$ to construct two scalar (pseudo-scalar) fields $\mathcal{E}$ and $\mathcal{B}$. The expression for pure $\mathcal{E}$ and $\mathcal{B}$ fields as defined in \citet{2007PhRvD..76d3001S, 2010PhRvD..82b3001Z} is: 
\begin{eqnarray}
\label{pseudo_E}
\mathcal{E}(\hat{n}) &=& -\frac{1}{2}[\bar{\eth}\bar{\eth}P_{+}(\hat{n}) + \eth \eth P_{-}(\hat{n})], \\
\label{pseudo_B}
\mathcal{B}(\hat{n}) &=& -\frac{1}{2i}[\bar{\eth}\bar{\eth}P_{+}(\hat{n}) - \eth \eth P_{-}(\hat{n})].
\end{eqnarray}
The pure $E$- and pure $B$- fields defined here are two mutually orthogonal scalar and pseudo-scalar fields. The $\mathcal{E}(\hat{n})$ and $\mathcal{B}(\hat{n})$ fields can be decomposed in terms of spherical harmonics as usual, with spherical harmonic coefficients $\mathcal{E}_{\ell m}$ and $\mathcal{B}_{\ell m}$. They can be computed from the pure fields as:
\begin{eqnarray}
 \label{Elm_2}
\mathcal{E}_{\ell m} &=& \int  \mathcal{E}(\hat{n})Y_{\ell m}^*(\hat{n})d\hat{n},\\
\mathcal{B}_{\ell m} &=& \int  \mathcal{B}(\hat{n})Y_{\ell m}^*(\hat{n})d\hat{n}.
\end{eqnarray}

This new pure $E$- and $B$-mode spherical harmonic coefficients are related to the full sky $E$-, and $B$-mode coefficients as \citep{zaldarriaga-b-mode}:
\begin{eqnarray}
\label{Elm_pseudo}
\mathcal{E}_{\ell m}=N_{\ell,2}E_{\ell m},\\
\mathcal{B}_{\ell m}=N_{\ell,2}B_{\ell m}.
\end{eqnarray}
For an incomplete sky observation defined by the window function $W(\hat{n})$, the partial-sky harmonic coefficients of $\mathcal E$- and $\mathcal B$-fields (with overhead tilde) are defined as \citep{efstathiou2004},
\begin{eqnarray}
\tilde{\mathcal E}_{\ell m} &=& -\frac{1}{2}\int d\hat{n} \bigg\{P_{+}(\hat n)\left[\bar\eth \bar\eth\left(W(\hat{n})Y_{\ell m}(\hat{n})\right)\right]^\ast   \nonumber \\
& & +P_-(\hat n)\left[\eth\eth\left(W(\hat{n})Y_{\ell m}(\hat{n})\right)\right]^\ast \bigg\}, \label{pureee}\\
\tilde{\mathcal B}_{\ell m} &=& -\frac{1}{2i}\int d\hat{n}\bigg\{P_+(\hat n)\left[\bar\eth\bar\eth\left(W(\hat{n})Y_{\ell m}(\hat{n})\right)\right]^\ast  \nonumber  \\
& & -P_-(\hat n)\left[\eth\eth\left(W(\hat{n})Y_{\ell m}(\hat{n})\right)\right]^\ast \bigg\} \label{purebb}.
\end{eqnarray}
These expressions can be expanded and simplified further for implementation and full expressions can be found in \citep{2016RAA....16...59W}. Once the coefficients $\tilde{\mathcal E}_{\ell m} $ and $\tilde{\mathcal B}_{\ell m}$ are derived, the scalar fields in our observation window $W(\hat{n})\mathcal{E}(\hat{n})$ and $W(\hat{n})\mathcal{B}(\hat{n})$ can be directly obtained by inverse harmonic transforms. 

The partial-sky pure harmonic coefficients $\tilde{\mathcal E}_{\ell m}$ and $\tilde{\mathcal B}_{\ell m}$ are related to the full-sky harmonic coefficients $E_{\ell m}$ and $B_{\ell m}$ as follows,
\begin{eqnarray}
\tilde{\mathcal E}_{\ell m} &=&\sum_{\ell'm'}[\mathcal{K}^{EE}_{\ell m,\ell'm'}E_{\ell'm'}+i\mathcal{K}^{EB}_{\ell m,\ell'm'}B_{\ell'm'}], \label{Elm}\\
\tilde{\mathcal B}_{\ell m} &=&\sum_{\ell'm'}[-i\mathcal{K}^{BE}_{\ell m,\ell'm'}E_{\ell 'm'}+\mathcal{K}^{BB}_{\ell m,\ell'm'}E_{\ell'm'}], \label{Blm}
\end{eqnarray}
where the pure field mixing kernels are denoted by $\mathcal{K}^{XY}_{\ell' m' \ell m}$. Their detailed expressions are given in \citep{2009PhRvD..79l3515G,ferte2013}. The cross mixing matrix for $EB$ or $BE$ for the pure method is orders of mgnitude smaller than that for the standard definition of Eq. (\ref{std_coupling_correalation}). This implies that the pure fields are nearly orthogonal with very small mixing between the two polarization modes.

\subsubsection{PCL estimator definition}
In this work, the PCL estimator is constructed with standard $E$-mode, and pure-$B$-mode definitions. The $T$-mode definition is unchanged. The cross-spectra estimators are defined as follows:
\begin{eqnarray}
\langle\hat \mathcal{C}_{\ell}^{TE}\rangle &\equiv& \frac{1}{2l+1}\sum_m \langle T^{(std)}_{\ell m} E^{(std)*}_{\ell m} \rangle, \\
\langle \hat \mathcal{C}_{\ell}^{T\mathcal{B}}\rangle &\equiv& \frac{1}{2l+1}\sum_m \langle T^{(std)}_{\ell m} \mathcal{B}_{\ell m}^* \rangle=N_{\ell ,2} C_{\ell}^{TB}, \\
\langle \hat \mathcal{C}_{\ell}^{E\mathcal{B}} \rangle &\equiv& \frac{1}{2l+1}\sum_m \langle E^{(std)}_{\ell m} \mathcal{B}_{\ell m}^* \rangle=N_{\ell ,2} C_{\ell}^{EB},
\end{eqnarray}
where $N_{\ell ,s} = \sqrt{(\ell + s)! / (\ell - s)!}$, and $\langle \cdots \rangle$ implies averaging over realizations. 

To reconstruct actual CMB cross power spectra, the relationship between partial sky cross spectra (denoted by overhead tilde) to the full sky spectra is necessary. For $TE$ and $TB$ part, 
\begin{eqnarray}
    \langle\tilde{\mathcal{C}}_\ell^{TE}\rangle &=& \sum_{\ell'} \left[ \mathcal{M}_{\ell \ell'}^{TE,TE}C^{TE}_{\ell'} + \mathcal{M}_{\ell \ell'}^{TE,TB}C^{TB}_{\ell'}\right], \label{eq:pureEClmix}\\
    \langle\tilde{\mathcal{C}}_\ell^{T\mathcal{B}}\rangle &=& \sum_{\ell'} \left[ \mathcal{M}_{\ell \ell'}^{TB,TE}C^{TE}_{\ell'} + \mathcal{M}_{\ell \ell'}^{TB,TB}C^{TB}_{\ell'}\right]. \label{eq:pureTBTEmix}
\end{eqnarray}
The $EB$ part is more complicated, and their relationship given by following expression,
\begin{eqnarray}
    \langle\tilde{\mathcal{C}}_\ell^{EE}\rangle &=& \sum_{\ell'} \left[ \mathcal{M}_{\ell \ell'}^{EE,EE}C^{EE}_{\ell'} + \mathcal{M}_{\ell \ell'}^{EE,BB}C^{BB}_{\ell'} + \mathcal{M}_{\ell \ell'}^{EE,EB}C^{EB}_{\ell'}\right] \label{eq:pureEClmix}\\
    \langle\tilde{\mathcal{C}}_\ell^{\mathcal{BB}}\rangle &=& \sum_{\ell'} \left[ \mathcal{M}_{\ell \ell'}^{BB,EE}C^{EE}_{\ell'} + \mathcal{M}_{\ell \ell'}^{BB,BB}C^{BB}_{\ell'} + \mathcal{M}_{\ell \ell'}^{BB,EB}C^{EB}_{\ell'} \right] \\
    \langle\tilde{\mathcal{C}}_\ell^{E\mathcal{B}}\rangle &=& \sum_{\ell'} \left[ \mathcal{M}_{\ell \ell'}^{EB,EE}C^{EE}_{\ell'} + \mathcal{M}_{\ell \ell'}^{EB,BB}C^{BB}_{\ell'} + \mathcal{M}_{\ell \ell'}^{EB,EB}C^{EB}_{\ell'}\right]
\end{eqnarray}
The detailed expression of the mixing matrices, $\mathcal{M}^{ru}_{\ell \ell'}$, can be derived from definition and are listed in appendix of \citep{ferte2013}.
The PCL estimator for this work has been implemented with the python package of NaMaster \footnote{https://github.com/LSSTDESC/NaMaster} \citep{Alonso2019}.

\subsection{Standard QML estimator} 
\label{sub:qml_estimator}
For CMB observations with any sky coverage \citet{2001PhRvD..64f3001T} defined the optimal QML estimator for temperature and polarization. In this section we will briefly review the QML estimator.
We define the input data vector, $\bm{x}$, consisting of the temperature, and the Stokes $Q$ and $U$ fields (with respect to a fixed coordinate system), at the $i^{\rm th}$ pixel as
 \begin{equation}       
\label{define_x}
\bm{x}_i=              
  \left(
\begin{array}{c}   
    \Delta T_i  \\  
    Q_i  \\  
    U_i  
  \end{array}     \right)             
  +
  \left(
\begin{array}{c}   
    n^T_i  \\  
    n^Q_i  \\  
    n^U_i  
  \end{array}     \right),
\end{equation}
where $n_i^X$ denotes the noise. The optimal quadratic estimate of the power spectrum, $y^{XY}_\ell$, is defined as \citep{2001PhRvD..64f3001T} :
\begin{equation} 	
 \label{QML_yrl}
	y^{XY}_\ell = \bm{x}^t_i \bm{E}_{\ell, ij}^{XY} \bm{x}_j - b^{XY}_\ell, \qquad X,Y \in \left[T,E,B\right],
\end{equation}
where $t$ indicates matrix transpose operation. Here $i$, $j$ are indices over pixels, and $\bm{E}_{\ell, ij}^{XY}$ is a $3 \times 3 $ matrix. The bias term, $b^{XY}_\ell$, corrects for the noise bias and is computed as ${\rm Tr}\left[\bs E^{XY}_\ell \bs N\right]$, assuming the noise to be uncorrelated between pixels. In these relations we have assumed the summation convention. The $\bm{E}^{XY}_\ell$ matrices are computed as:
\begin{eqnarray}	
	\bm{E}^{XY}_\ell= \frac{1}{2} \bm{C}^{-1} \frac{ \partial \bm{C} }{ \partial C^{XY}_\ell } \bm{C}^{-1}
	\label{QML_Erl}
\end{eqnarray}
with the covariance matrix of $\bm{x}$ denoted by $\bm{C}$. The detailed expressions for the covariance matrix can be found in \citet{2001PhRvD..64f3001T}.

The $y^{XY}_\ell$ gives unbiased estimate the actual power spectra $C^{XY}_\ell$. Using Eq. (\ref{QML_yrl}) and Eq. (\ref{QML_Erl}) we can get the expectation value of $y^{XY}_\ell$ as:
\begin{equation}	
	\langle y^{XY}_\ell \rangle = F^{XYPQ}_{\ell \ell'}C^{PQ}_{\ell'},
	\label{QML_brave_yrl}
\end{equation}
where $X,Y,P,Q \in [T, E, B]$, and we have used the Fisher matrix defined as:
\begin{equation}
    F^{XYPQ}_{\ell \ell'}=\frac{1}{2}{\rm Tr}\left[\frac{\partial \bs C} {\partial C^{XY}_{\ell'}} \bs C^{-1} \frac{\partial \bs C}{\partial C^{PQ}_{\ell}} \bs C^{-1}\right]
     \label{QML_Fisher_Matrix}.
\end{equation}

When $\bs F$, is invertible, one can define unbiased estimates of the true power spectra via
\begin{equation}
    \hat{C}^{XY}_\ell=F^{-1}\bs y^{XY}
    \label{QML_true_PS}.
\end{equation}	
The covariance matrix of $y^{XY}_\ell$ is then given by:
\begin{eqnarray}	
	\langle y^{XY}_\ell y^{PQ}_{\ell'}\rangle - \langle y^{XY}_\ell\rangle \langle y^{PQ}_{\ell'} \rangle \equiv F^{XYPQ}_{\ell \ell'} = 2{\rm Tr}\left[\bs C \bs E^{XY}_\ell \bs C \bs E^{PQ}_{\ell'}\right],
	\label{QML_covariance_matrix1}
\end{eqnarray}
with $\bs F^{XYPQ}$ being the Fisher matrix. Hence, the covariance matrix of the actual power spectra estimates of Eq.  (\ref{QML_true_PS}) is:
\begin{equation}	
	\langle \Delta \hat{C}_\ell \Delta \hat{C}_{\ell'} \rangle = \bs F^{-1}.
	\label{QML_covariance_matrix2}
\end{equation}
The QML estimators for this work have been implemented with the xQML\footnote{https://gitlab.in2p3.fr/xQML/xQML} python package \citep{Vanneste2018}.



\subsection{QML-SZ estimator}
\label{sub:qml_sz_estimator}
The standard QML estimator, for CMB $TQU$ maps, described before is optimal but it is computationally prohibitive at high resolutions. We attempt to reduce the size of the computation problem by essentially reducing the estimation of each cross spectra as its own scalar problem. We have previously demonstrated this method for $B$-mode auto spectrum in \citet{chen2021fast}. 

We use the pure method definitions of simple harmonic coefficients to compute pure-$E$- and pure-$B$-mode fields. With this definition any correlations due to $E$-to-$B$ or $B$-to-$E$ leakages should be suppressed by few orders of magnitude. We would then treat the cross-spectra computation as a scalar problem with the scalar temperature-only QML method. We will outline this `scalar' treatment of the cross-spectra below.

Let $d^X_{i}$ denote the $i^{th}$ pixel value in the scalar map $s^X_{i} + n^X_{i}$, where $s^X_{i}$ is the signal and $n^X_{i}$ is the noise in the individual pixel. The covariance matrix $\bs C^{XY}$ of observation $d^X_{i}$ and $d^Y_{i}$, with $X,Y \in [ T, \mathcal E, \mathcal B]$, and $X \ne Y$, can be written as:
\begin{eqnarray} 
\label{QML-SZ_matrix_C}
C^{XY}_{ij}=\langle \bs{d}^X (\bs{d}^Y)^t \rangle =  \sum_{\ell}\frac{2 \ell + 1}{4 \pi} C_{\ell}^{XY} P_{\ell}(z)+N^{XY}_{ij} 
\end{eqnarray}
where $C_{\ell}^{XY}$ is the cross-spectrum corresponding to the scalar signal maps $s^X(\hat{n})$ and $s^Y(\hat{n})$. \textbf{$P_\ell$ denotes a Legendre polynomial and $z$ is cosine of the angle between the two pixels under consideration.} Here $N^{XY}_{s,ij}$ is the noise covariance matrix. 

We can define the quadratic estimator with `scalar' approximation as:
\begin{equation} 	
 \label{QML_SZ_yl}
	y^{XY}_{\ell} =(\bs{d}^X)^t  \bs{E}^{XY}_{\ell} \bs{d}^Y - b^{XY}_{\ell}.
\end{equation}
\textbf{The matrices $\bs{E}^{XY}_{\ell}$ have a similar form to equation in Sec.B of \citet{Vanneste2018}:}
\begin{equation}	
 \label{QML_SZ_Erl}
    \bs{E}^{XY}_{\ell} = \frac{1}{2}  (\bs{C}^{XX})^{-1} 
    \frac{\partial \bs{C}^{XY}}{\partial C^{XY}_{\ell}}
    (\bs{C}^{YY})^{-1}.
\end{equation}
where $\frac{\partial \bs{C}^{XY}}{\partial C^{XY}_{\ell}} =C^{XX}\bs{E}_\ell C^{YY}+C^{XY}\bs{E}^t_\ell C^{XY}$
Likewise, the \textbf{mode-mixing} matrix, $F_{\ell \ell'}$, expression becomes:
\begin{equation} 	
\label{QML_SZ_Fllp}
    F_{\ell \ell'} = \frac{1}{2}Tr \left[ (\bs{C}^{XX})^{-1} \frac{\partial \bs{C}^{XY}}{\partial C^{XY}_{\ell}}  (\bs{C}^{YY})^{-1}\frac{\partial \bs{C}^{XY}}{\partial C^{XY}_{\ell'}} \right].
\end{equation}

Finally, the QML estimator for the cross power spectrum $\hat C_{\ell}^{XY}$ is given by
\begin{equation} 	
 \label{QML_SZ_true_PS}
	\hat{C}_{\ell}^{XY}=\left(F_{\ell \ell '}\right)^{-1} y^{XY}_\ell.
\end{equation}

When computing the $\mathcal E$- or $\mathcal B$-fields, we must use a proper sky apodization to avoid numerical divergences in the calculation of the window function derivatives. \citet{2016RAA....16...59W} and \citet{2011A&A...531A..32K} have shown that a Gaussian smoothing kernel induces very small leakage in the final map. In this work, we will use Gaussian apodization to obtain the $\mathcal E$- and $\mathcal B$-maps for the QML-SZ method. For the $i^{\rm th}$ pixel in the region allowed by the binary mask, the apodized window is defined as:

\begin{numcases} {W_i=} 
         \frac{1}{2} +\frac{1}{2}{\rm erf}\left(\frac{\delta_i-\frac{\delta_c}{2}}{\sqrt{2}\sigma}\right), & $\delta_i < \delta_c$  \nonumber \\
         1, & $\delta_i > \delta_c$ \label{QML_SZ_Gaussian_window_function}
\end{numcases}
where $\delta_i$ is the closest distance between the $i^{\rm th}$ observed pixel from the boundary of the allowed region,  $\sigma={\rm FWHM}/\sqrt{8\ln 2}$ with FWHM denoting the full width at half maximum of the Gaussian kernel, and $\delta_c$ is the apodization length which acts as an additional adjustable parameter.

The steps to estimate the cross spectrum with the QML-SZ estimator can be summarized as follows: First, for the given observed $Q$ and $U$ polarization maps, we construct a partial-sky pure-$E$-map, $\mathcal{E}(\hat{n})$ and pure-$B$-map, $\mathcal{B}(\hat{n})$, using a Gaussian apodized window function. For T map, we just use observed T map directly. Then, we use the two maps and the corresponding fiducial cross spectrum as input, to estimate the cross spectrum $C_\ell^{XY}$. In comparison to the standard QML estimator, our goal with the QML-SZ method is to simplify the calculation, without compromising significantly on the accuracy or error bars. We implement the QML-SZ estimator with the modified xQML python package.
\section{SIMULATION SETUP} 
\label{sec:SIMULATION_SETUP}
\begin{figure}[t]
\centering
\includegraphics[width=0.48\textwidth]{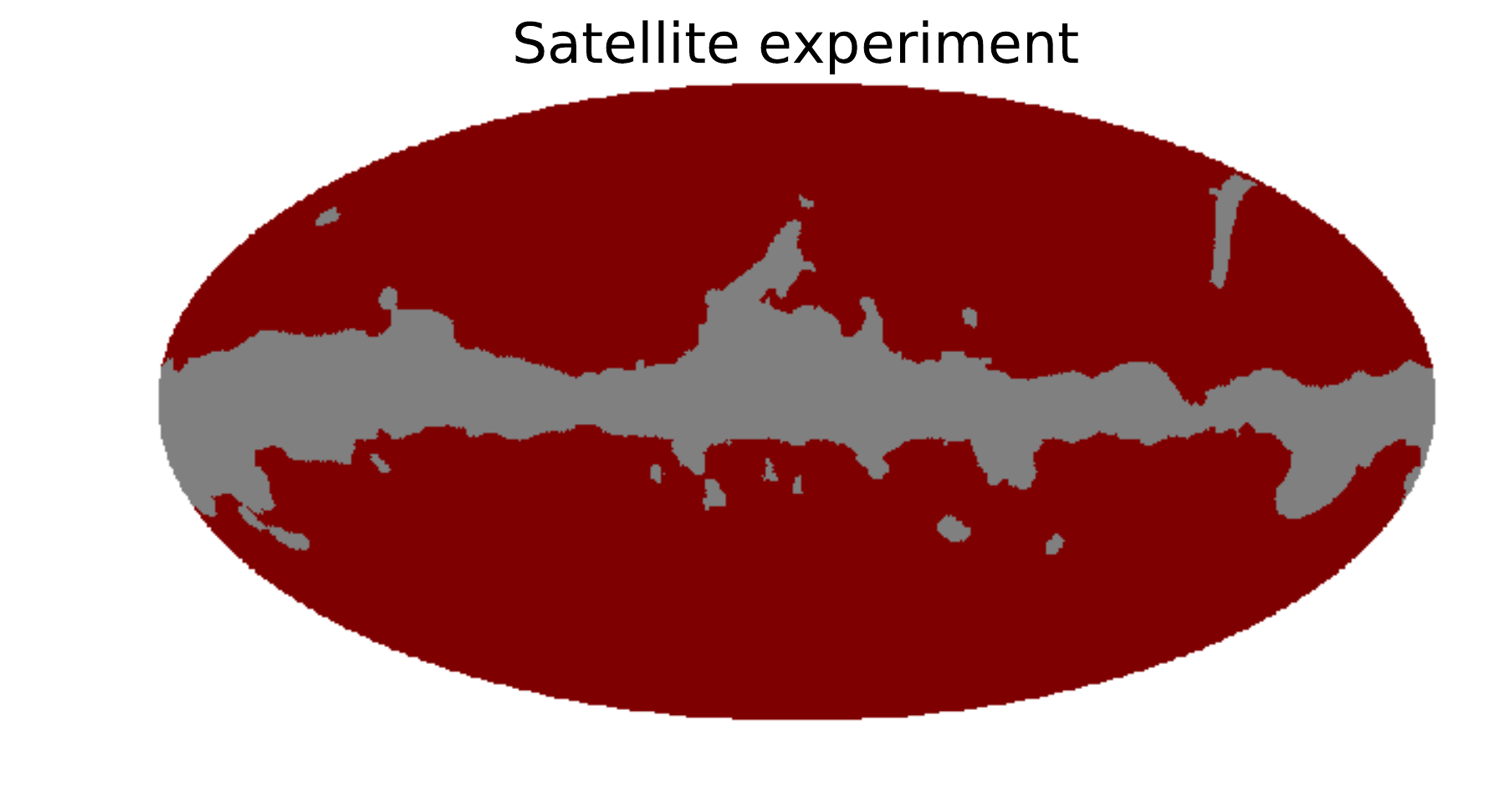}
\includegraphics[width=0.38\textwidth]{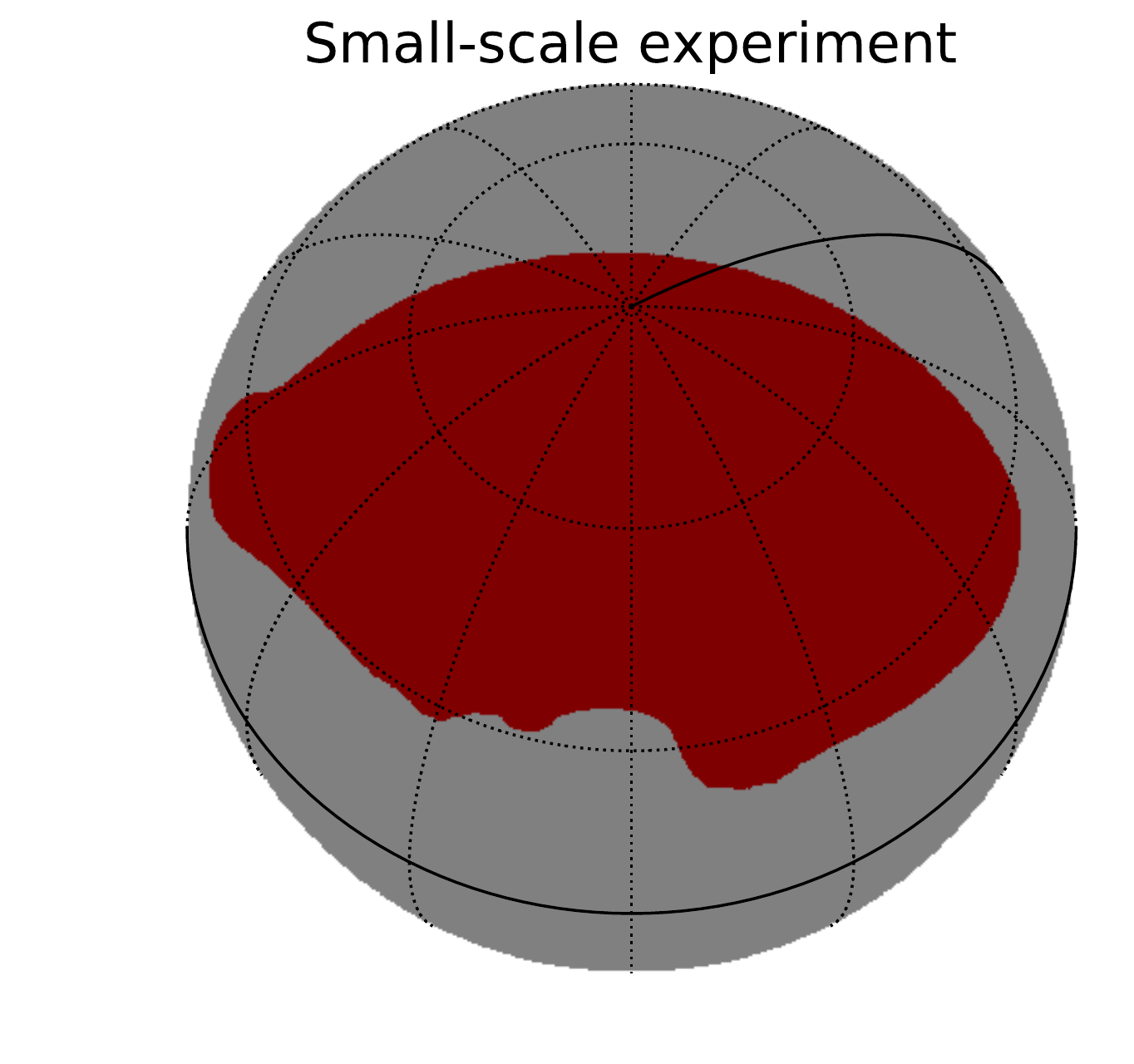}
\caption{Binary masks showing the observed sky patch for a space-based experiment (top) and for the ground-based experiment (bottom) as considered in this work. The red area is the observed area and the gray area is the mask area. The plots are in Galactic coordinate system. The sky fractions are $78.8\%$ and $15.1\%$, respectively.} 
\label{sky_coverage} 
\end{figure}

In this section, we will outline the simulation pipeline used in this work. 
Here, we consider two cases of future CMB polarization experiments: a space-based, and a ground-based CMB polarization experiment. 
For the space-based experiment case, we consider the 2018 Planck common polarization mask, which masks the galactic foregrounds and the point sources resolved in Planck maps. We fill-in all the point source smaller than $5^\circ$, as well as the extended source masking at high galactic latitudes ($|b|>45^\circ$) by using HEALPix \texttt{process\_mask} subroutine \footnote{http://healpix.sourceforge.net}. So we obtain a $\sim 78\%$ sky coverage patch for the space-based CMB experiment showed in the upper panel of figure \ref{sky_coverage}. Second, we consider the AliCPT-1 experiment \citep{Hong2017, 2021arXiv210109608S}, a CMB experiment in the northern hemisphere, to represent of a ground-based CMB experiment. The observed area covers $\sim 15.1 \%$ 
of the full sky and the binary mask of it be showed in the lower panel of Fig. \ref{sky_coverage}. 

The observed data is assumed to consist of two parts: the CMB signal, and the instrumental noise. The input power spectra for CMB maps are computed with CAMB\footnote{http://camb.info} \citep{Lewis:1999}, using the 2018 Planck cosmological parameters as given by \citet{Planck2018VI}, with lensing and the tensor-to-scalar ratio $r=0.05$. The CMB maps are produced using the \texttt{synfast} subroutine of HEALPix\footnote{http://healpix.sourceforge.net} at $\texttt{NSIDE}=512$ with $\ell_{\rm max}=1024$. For the noise map, we assumed a Gaussian homogeneous noise model with the noise RMS set to $3\mu K$-arcmin, which is a typical value for the next generation of space-based experiments \citep{litebird,pico,core,zhao2011,huang2015}. This equates to a white noise level of $0.44\mu k$-pixel at at $\texttt{NSIDE}=512$. Finally, adding the CMB signal map and noise map together we obtain the simulated observation. Depending on the method we further pre-process these maps, as detailed below.

\subsection{Space-based experiment}
\label{sub:space_based_simulation}
 After obtaining the observed maps, the subsequent pre-processing steps are different for the three estimators:

\textit{PCL estimator}: We smooth the masked $TQU$ map with a Gaussian smoothing with FWHM$=20'$ first.  We apodize our observation window with a `C2' (cosine) apodization function for the PCL estimator to reduce mode mixing and leakage in this work. The weight in the $i^{\rm th}$ pixel is given as \citep{Alonso2019}:

    \begin{numcases} {W_i =}
    \label{C2}
    \frac{1}{2}\left[1 - \cos (\pi \delta^r_i)\right] & $\delta^r_i < 1$ \nonumber\\
    1 & ${\rm otherwise}$,
    \end{numcases}
where $\delta^r_i = \sqrt{(1-\cos \delta_i)/(1-\cos \delta_c)}$. For space-based experiment case $\delta^r_i$ be set to $6^\circ$. 

\textit{Standard QML estimator}: Downgrading high resolution maps to low resolution maps is the key of standard QML estimator. 
To extract the large scale information, we degrade the resolution to HEALPix  \texttt{NSIDE}=16 with the following smoothing:
\begin{numcases} {f(\ell)=} 
         1, &$\ell \leq N_{side}$\\
         \frac{1}{2} (1+\cos(\frac{(l-N_{side})\pi}{2N_{side}})). & $N_{side} < \ell \leq 3N_{side}$  \nonumber \\
         0, & $\ell > 3N_{side}$ \label{QML_SZ_Gaussian_window_function}
\end{numcases}
We smooth the $a_{\ell m}$s, obtained from the input maps at $\texttt{NSIDE}=512$, with the smoothing function of equation (\ref{QML_SZ_Gaussian_window_function}). We then obtain the smoothed map by performing inverse spherical harmonic transforms.

\textit{QML-SZ estimator}: In the QML-SZ estimator, the power spectrum of the pure-$E$ map $\mathcal{E}(\hat{n})$  and pure-$B$ map $\mathcal{B}(\hat{n})$  is blue. We smooth the observed maps with FWHM=$8^{\circ}$ to suppress higher multipoles, and then mask them. We then use the SZ-method to derive the pure-$E$ map $\mathcal{E}(\hat{n})$ and pure-$B$ map $\mathcal{B}(\hat{n})$, using a Gaussian apodized mask with  $\sigma = 10^{-6}$, $\delta_c =1^{\circ}$. Finally, using the \texttt{ud\_grade} subroutine of HEALPix to downgrade smoothed mask, $T$ map, pure-$E$-mode map, and pure-$B$-mode map to \texttt{NSIDE}=16.  In the smoothed mask we set all pixels with values $<0.99$ to zero. 
\begin{figure*}[th]
\centering
\includegraphics[width=\textwidth]{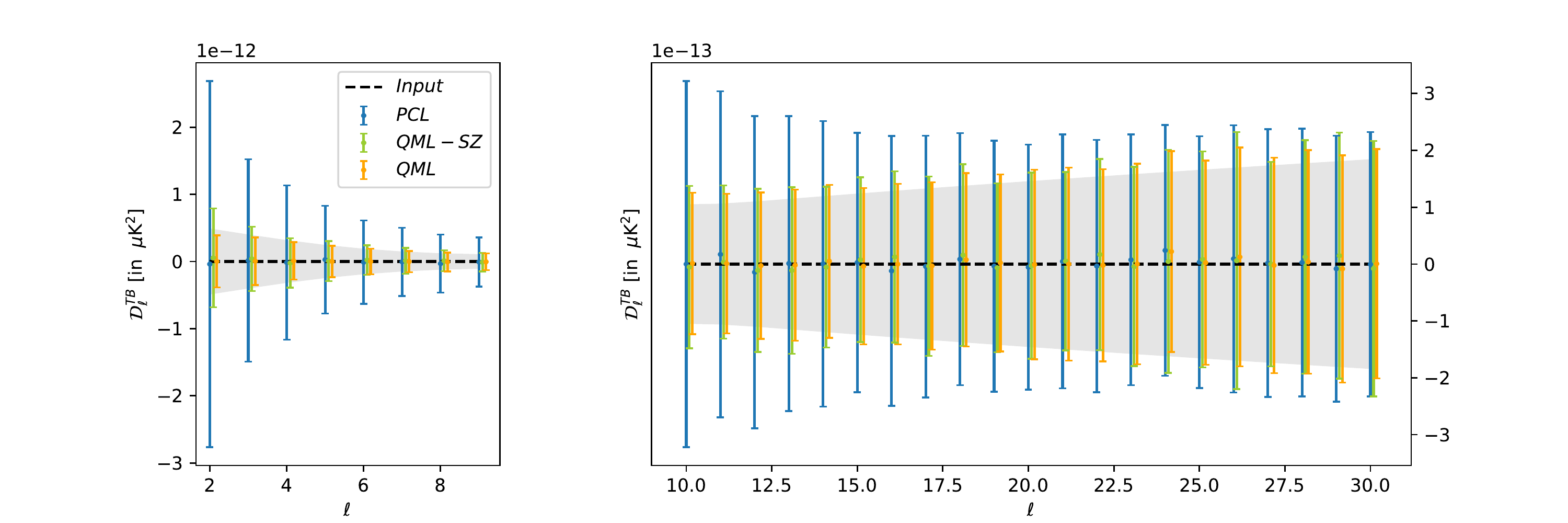}
\includegraphics[width=\textwidth]{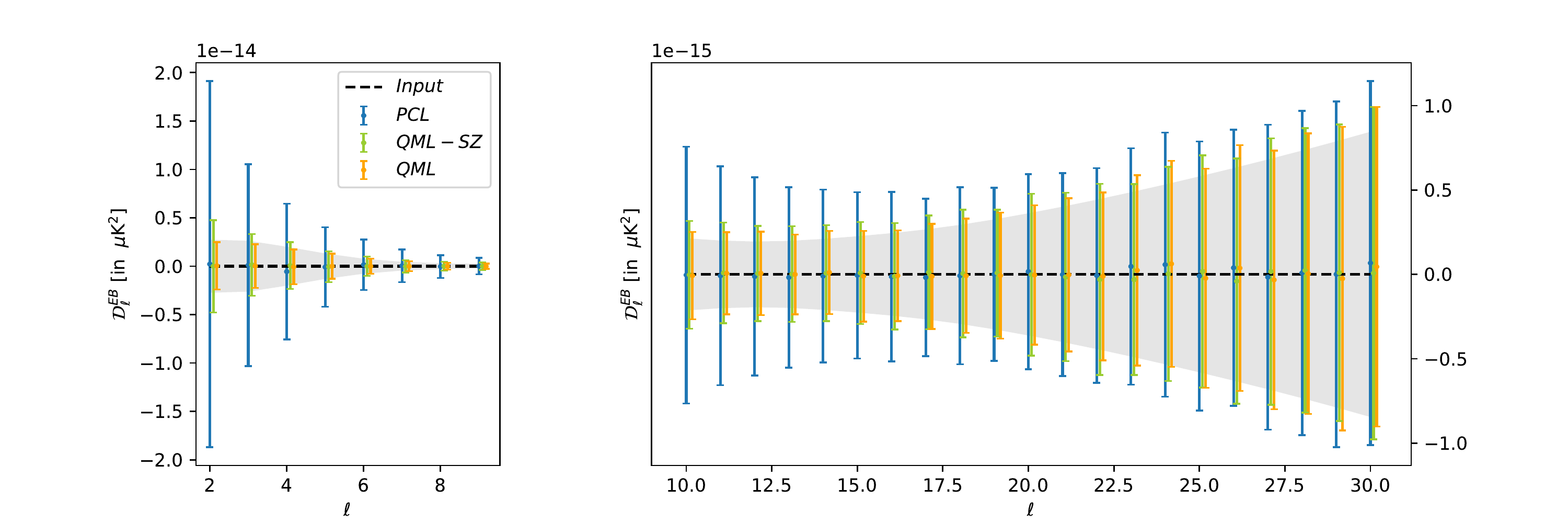}

\caption{Plot of the results for $TB$ and $EB$ power spectrum estimates for the realistic space-based CMB experiment, upper and lower panels are $TB$ and $EB$ respectively . The observed sky is simulated at \texttt{NSIDE}=512 with $\ell_{\rm max}$=1024 and the noise level is set to 3$\mu K$-arcmin. The input power spectra are shown with the black curve. The classic QML method results are shown with orange line, QML-SZ method results with green line. These results are computed at \texttt{NSIDE}=16. We also show PCL estimator results, obtained with NaMaster, with $\delta_c = 6^\circ$ C2 apodization, with blue line. The gray region denotes the analytical approximation of the error bounds. The data points are the mean of 1000 estimates, and the error bars are given by the standard deviation of the estimates.}
\label{fig:Planck_TB_EB}
\end{figure*}

\subsection{Ground-based experiment}
\label{sub:ground_based_simulation}
For the ground-based experiment, the coadded maps are multiplied by the AliCPT binary mask (see Fig.\ref{sky_coverage}) to keep only the fraction of sky observed in the considered experiment. Subsequent processing for hybrid PCL estimator and Standard QML estimator are similar to space-based experiment case, we just modify values of some parameters. For hybrid PCL estimator,  $\delta^r_i$ in equation (\ref{C2}) set to $10^\circ$. For standard QML estimator we set the target resolution $\texttt{NSIDE}=32$. But for QML-SZ estimator, we use different way to
deal with 'observed' CMB maps and noise maps. First, we use the SZ-method to obtain the pure $E$-mode and $B$-mode maps from smoothed $Q$ and $U$ maps,  using a Gaussian apodized mask with  $\sigma = 10^{-4}$, $\delta_c =0.5^{\circ}$.  We obtain the spherical harmonic coefficients of the maps at $\texttt{NSIDE}=512$, and set all harmonic coefficients to zero above $\ell_{\rm max}$. We use these $a_{\ell m}$s with this cut-off to reconstruct the map at $\texttt{NSIDE}=512$, but without the information above $\ell_{\rm max}$. This methodology is applied to the $T$ map, $\mathcal{E}$ and $\mathcal{B}$ maps at $\texttt{NSIDE}=512$ with $\ell_{\rm max}=192$. Then we downgrade these maps to the targeted $\texttt{NSIDE}=32$. 

\section{REALISTIC EXAMPLES} 
\label{sec:realistic_examples}
In this section, we discuss the results of the above estimators applied to estimate cross-correlation power spectra at the space-based experiment case in subsection \ref{sub:space_example} and the ground-based experiment case in subsection \ref{sub:ground_example}. We compare the computational requirements for these three methods in subsection \ref{sub:computation}. The QML-SZ estimator can be used to estimate $TE$ power spectrum. However, since the $TE$ power spectrum does not involve the $E$-$B$ leakage problem, and 
it is signal dominated, hybrid PCL estimator is adequate and we don't need to reconstruct $TE$ power spectrum using QML-SZ estimator. We only show our results of $TB$ and $EB$ spectrum estimates here. The $TE$ spectrum result is  shown and discussed in Appendix \ref{apdx:TE}. In this work, our power spectra estimates for any estimator is a mean of 1000 random simulations, and the errors are computed as the standard deviation of the samples.

\subsection{Space-based experiment} 
\label{sub:space_example}
One major advantage of space-based experiments is the ability to observe the full sky and therefore making measurements of the lowest multipoles of power spectra. We simply use our estimators on the pre-processed maps to obtain the $TB$ and $EB$ cross spectra.

The results for space-based experiment are shown in Fig.\ref{fig:Planck_TB_EB}. Considering that there is a significant difference in the order of magnitude on the error bars for $\ell < 10$ and for $\ell \ge 10$, the total range is split into two figures.
We find that for both cross spectra, all three estimators can get unbiased 
estimates of input power spectrum but all the QML estimators have smaller error bars in the entire multipole range. We also notice that while the standard QML method has nearly-optimal error bars throughout the entire multipole range, the QML-SZ method has sub-optimal error bars for the lowest multipoles. The error bars for $\ell \le 5$ show a significant increase for the QML-SZ estimator. This behaviour \textbf{is} caused by the blue input power spectra for the QML-SZ method due to the $N_{\ell,2}$ or $N^2_{\ell,2}$ factors in the cross spectra. Downgrading the map using \texttt{ud\_grade} there is some power leakage from high multipoles to low multipoles, leading to increase in uncertainty at low multipoles, as shown in \citep{chen2021fast}.


\begin{figure}[ht]
\centering
\includegraphics[width=0.23\textwidth]{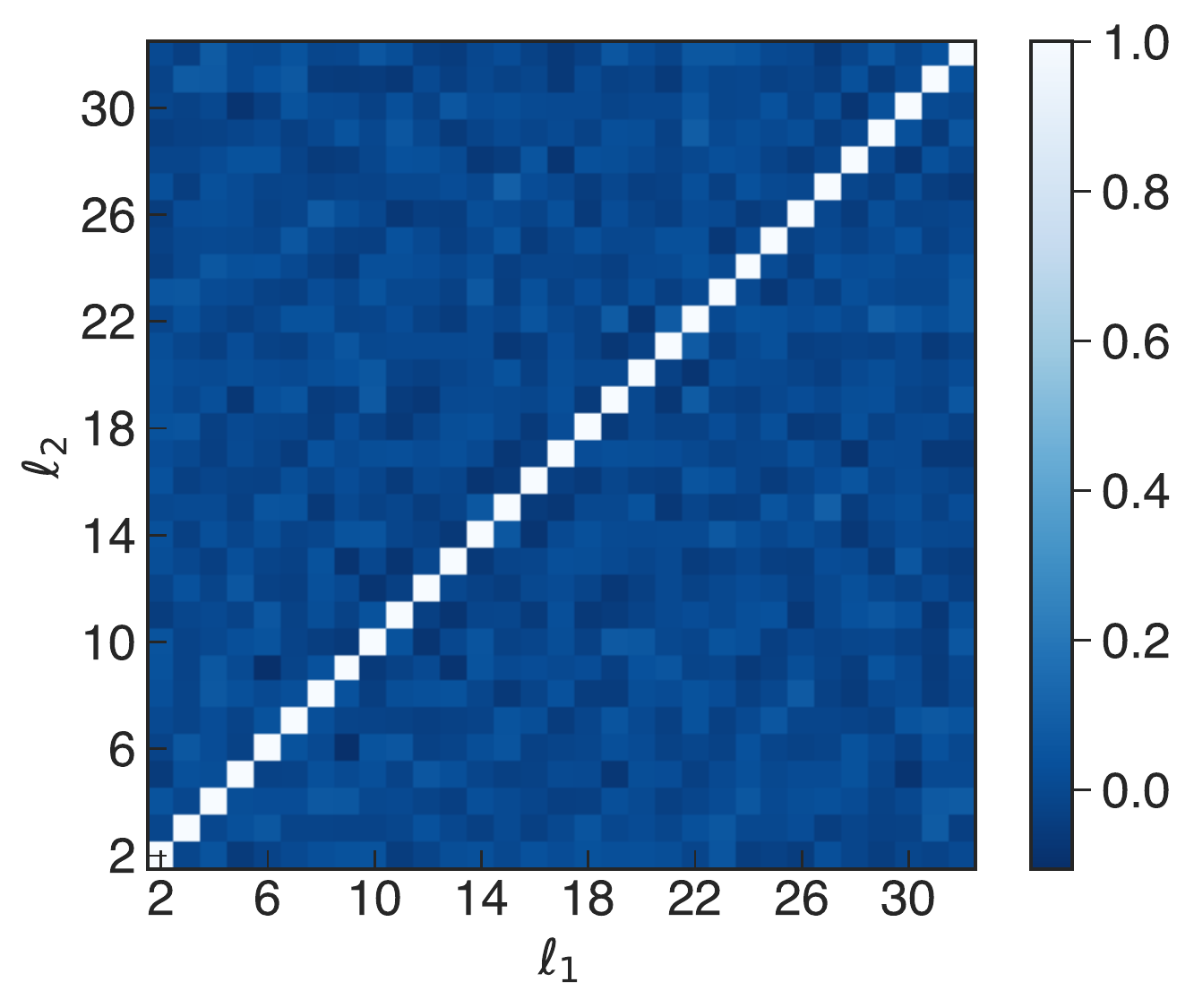}
\includegraphics[width=0.23\textwidth]{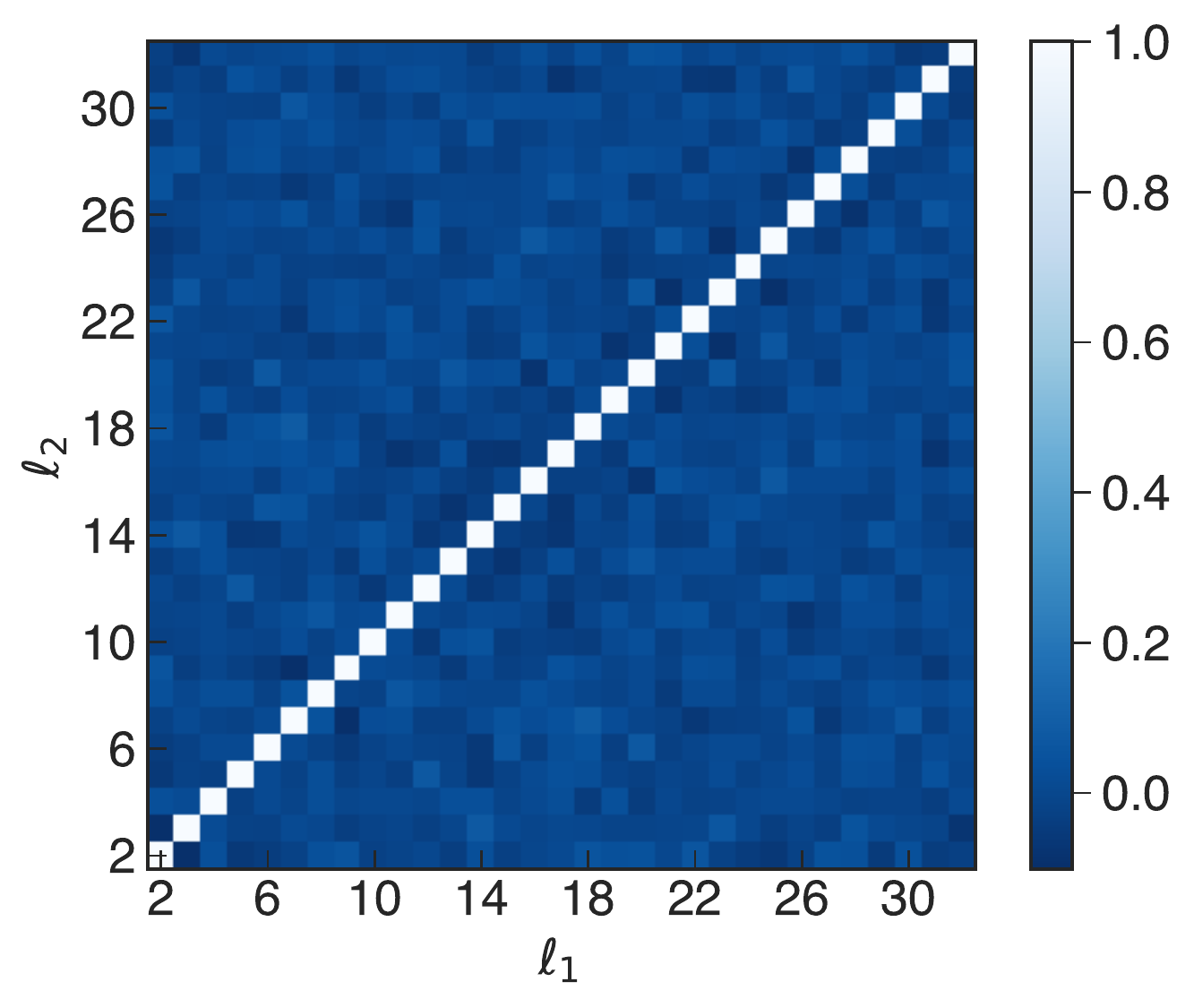}\\
\includegraphics[width=0.23\textwidth]{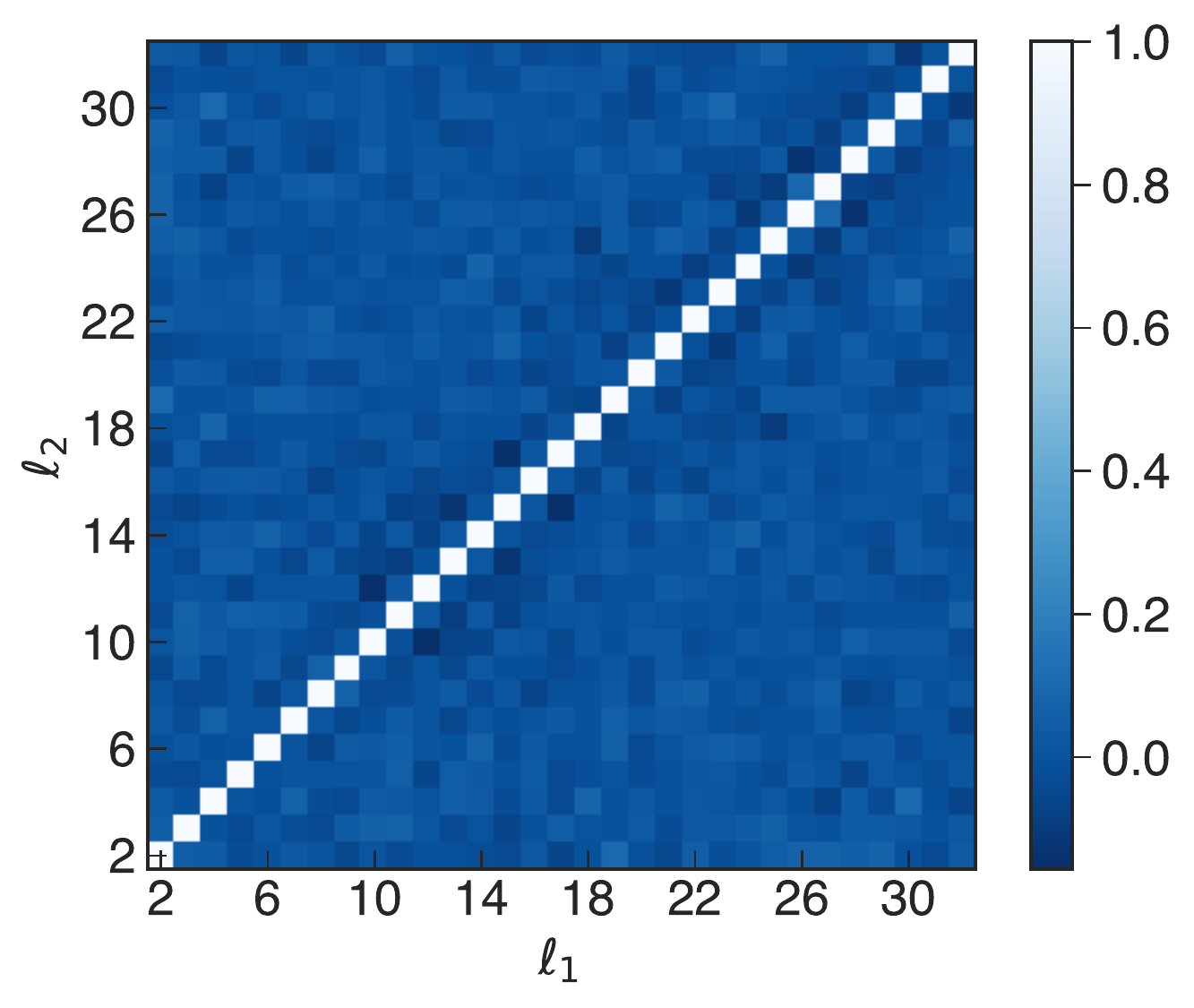}
\includegraphics[width=0.23\textwidth]{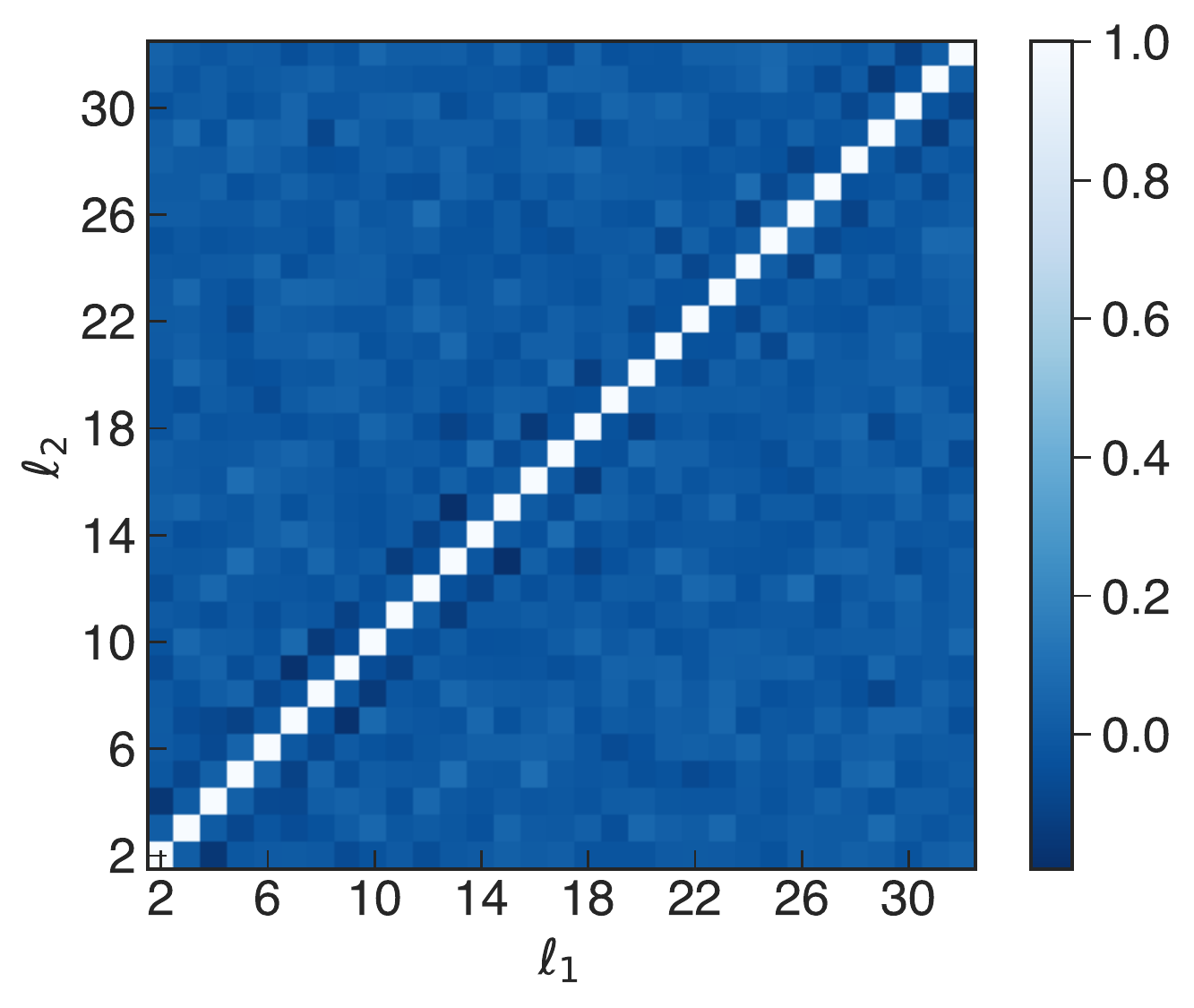}
\caption{Normalized covariance matrices $\mathcal{C}_{\ell, \ell'} = {\rm cov}(\hat{C}_\ell, \hat{C}_{\ell'}) /\sqrt{{\rm var}(\hat{C}_\ell){\rm var}(\hat{C}_{\ell'})}$ of the two QML methods for the space-based experiment with homogeneous noise. The matrices are obtained from estimates of 1000  simulations for classic QML estimator (upper panel) and for QML-SZ estimator (lower panel). The left diagrams show the covariance matrices of $TB$ mode, the right diagrams show equivalent plots of $EB$ mode. }
\label{fig:planck_cov}
\end{figure}
We note that the QML estimator uses binary mask, while the QML-SZ method uses apodized mask. This apodization reduces the effective $f_{\rm sky}$ for the QML-SZ method. While the performance of the QML-SZ estimator is not as good as the standard QML method, it is still a fast and reliable solution for power spectrum estimation, except for the lowest few multipoles. 
\begin{figure}[th]
\centering
\includegraphics[width=0.45\textwidth]{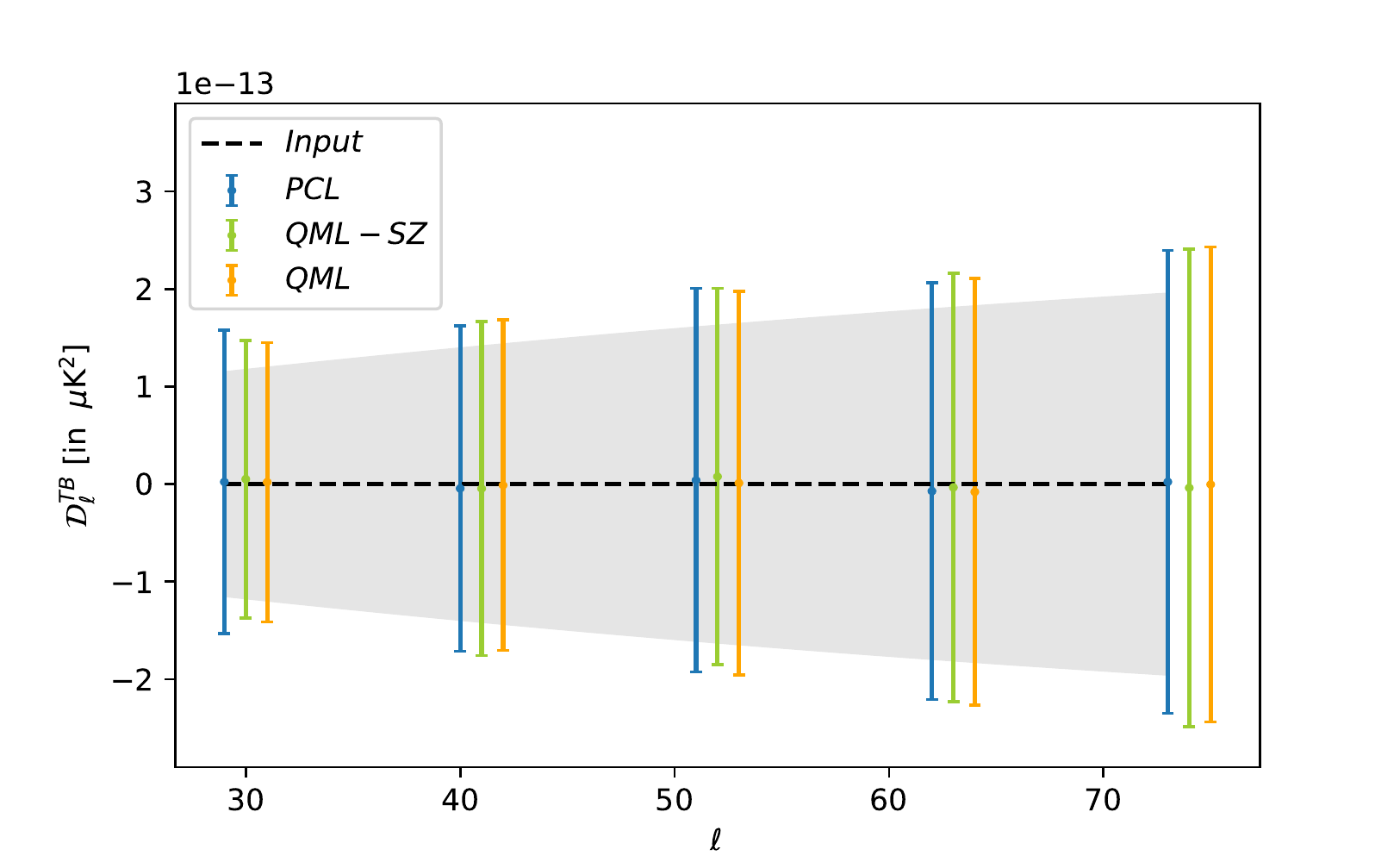}
\includegraphics[width=0.45\textwidth]{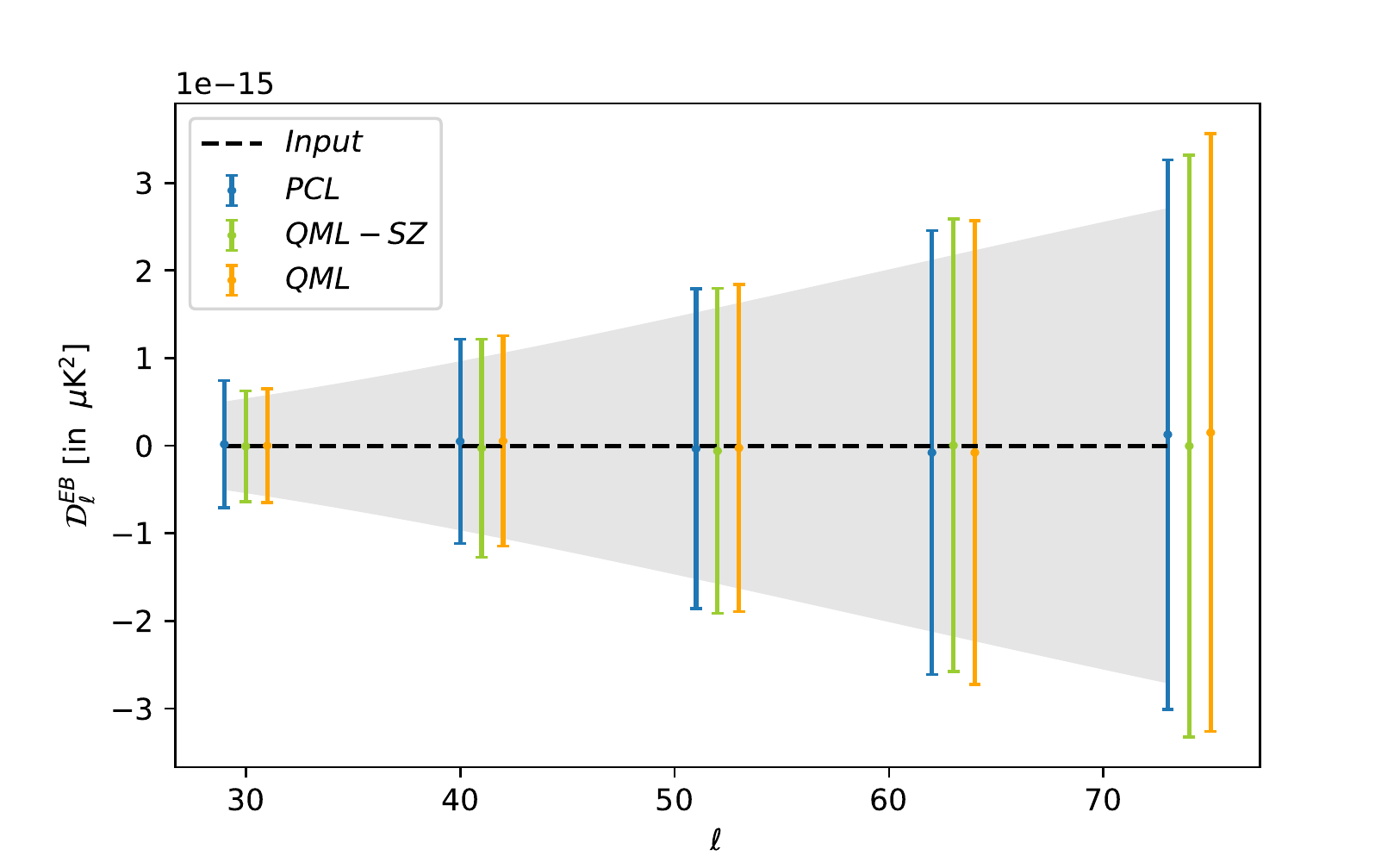}
\caption{Plot of the results of $TB$ and $EB$-mode power spectrum estimates for realistic ground-based CMB experiment with homogeneous noise, the upper panel and the lower panel are $TB$ and $EB$ respectively. The observed sky with 3 $\mu$K-arcmin noise, is simulated at \texttt{NSIDE}=512 with $\ell_{\rm max}$=1024. The input $B$-mode power spectrum is shown with the dotted, black curve. The classic QML method results are computed at \texttt{NSIDE}=32 with $\ell_{\rm max}$=96 (
orange), QML-SZ method results are computed at \texttt{NSIDE}=32 with $\ell_{\rm max}$=96 (green). We also show PCL estimator results, obtained with NaMaster, using $\delta_c = 10^\circ$ for $r=0.05$(blue). The gray region denotes the optimal error bounds. The data points are mean of 1000 estimates and the error bar is given by the standard deviation of the estimators.} 
\label{fig:realex_10} 
\end{figure}

As we are working on a partial sky the different multipoles are coupled.  We can use the normalized covariance matrices to quantify the coupling between different multipoles and it defined as
\begin{equation}
    \mathcal{C}_{\ell \ell'} = \frac{{\rm cov}\left(\hat{C}_\ell, \hat{C}_{\ell'}\right)}{\sqrt{{\rm var}(\hat{C}_\ell){\rm var}(\hat{C}_{\ell'})}},
    \label{eq:cov_mat}
\end{equation}
and the results of space-based experiment case are shown in Fig.\ref{fig:planck_cov}. For the case considered here, all the covariance matrices being approximately diagonal which means the power spectra estimates only weakly coupled among different multipoles.

\subsection{Ground-based experiment} 
\label{sub:ground_example}

In ground-based experiments we observe only a small fraction of the sky but with high sensitivity. This means we can choose larger \texttt{NSIDE} and extend the $\ell_{max}$ to higher multipoles. Here, we use AliCPT binary mask with $f_{sky}\sim 15.1\%$ to calculate the final results, and the estimated power spectra are binned with a band width of $n_{bin}$ = 11. 

The results of reconstruct cross-correlation power spectra are shown in Fig. \ref{fig:realex_10}. We find that all the methods discussed here can give unbiased estimates of both $TB$ and $EB$ band powers. \textbf{When we focus on the error bars, we find that all three methods have near-optimal error bars in the entire multipole range. As the performance of the PCL estimator is as good as the QML methods}, considering it is faster than QML-based estimators, it seems that we do not need to apply QML-based estimators to reconstruct cross-correlation power spectrum for ground-based experiment case.

\begin{figure}[h]
\centering
\includegraphics[width=0.23\textwidth]{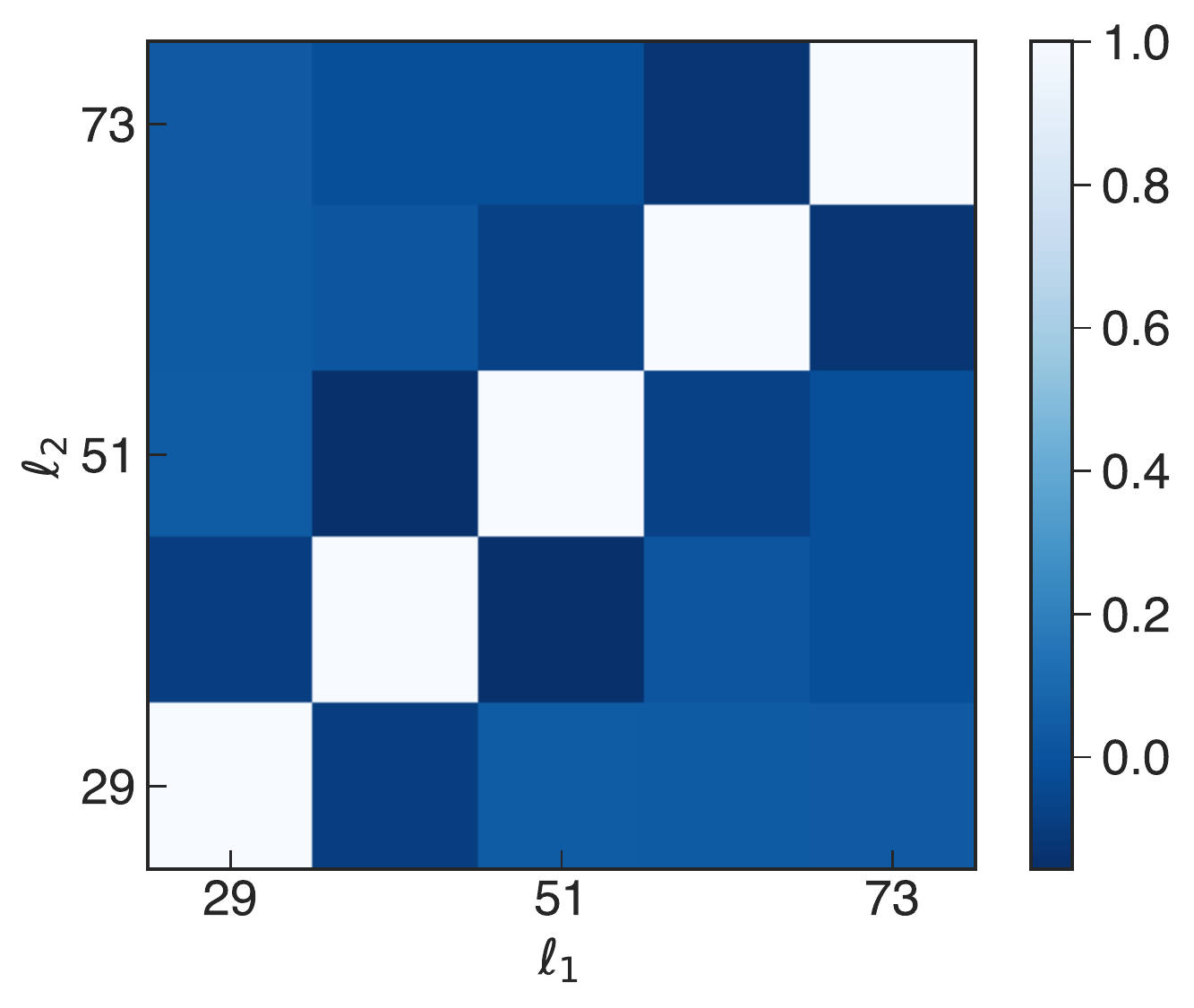}
\includegraphics[width=0.23\textwidth]{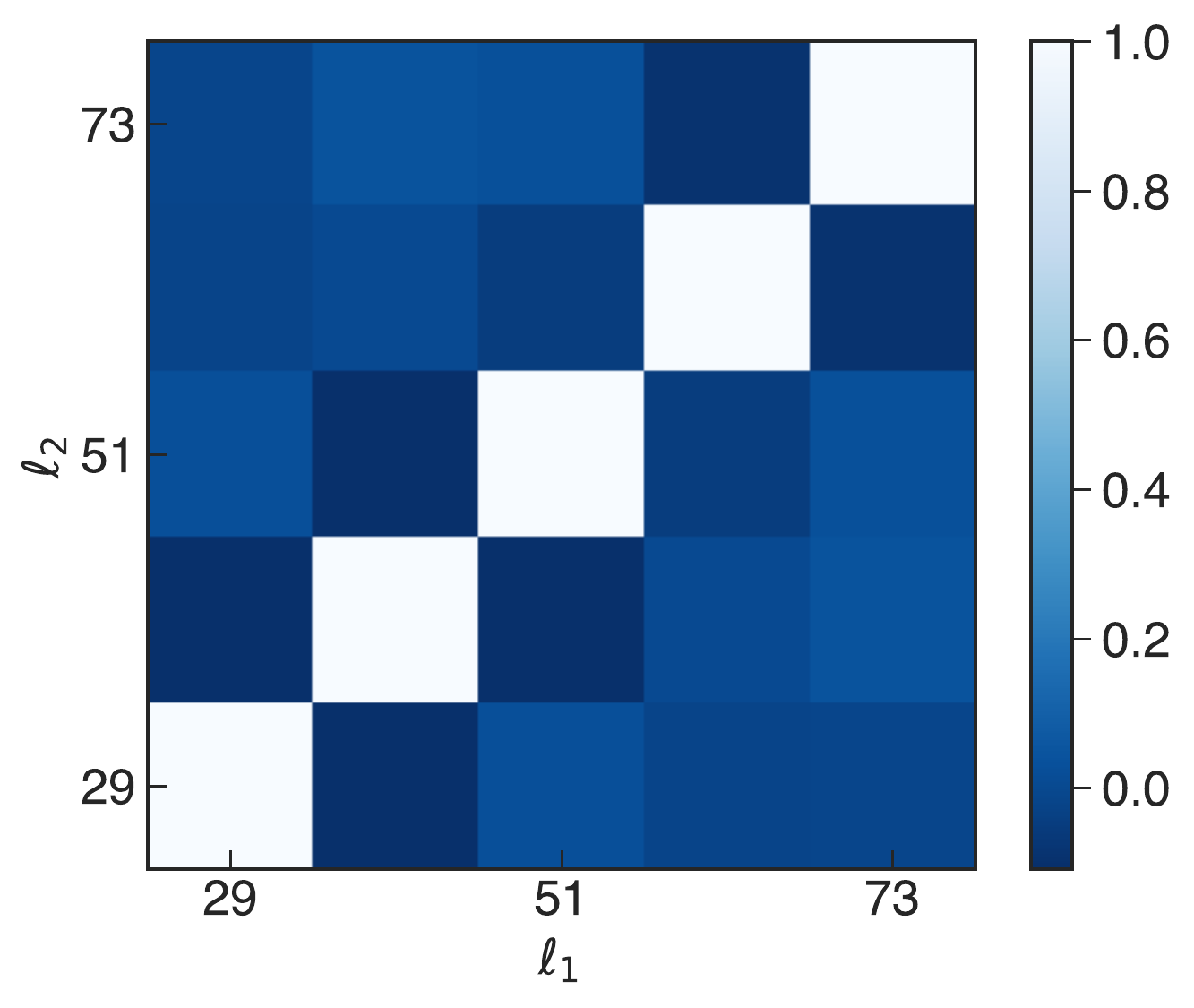}\\
\includegraphics[width=0.23\textwidth]{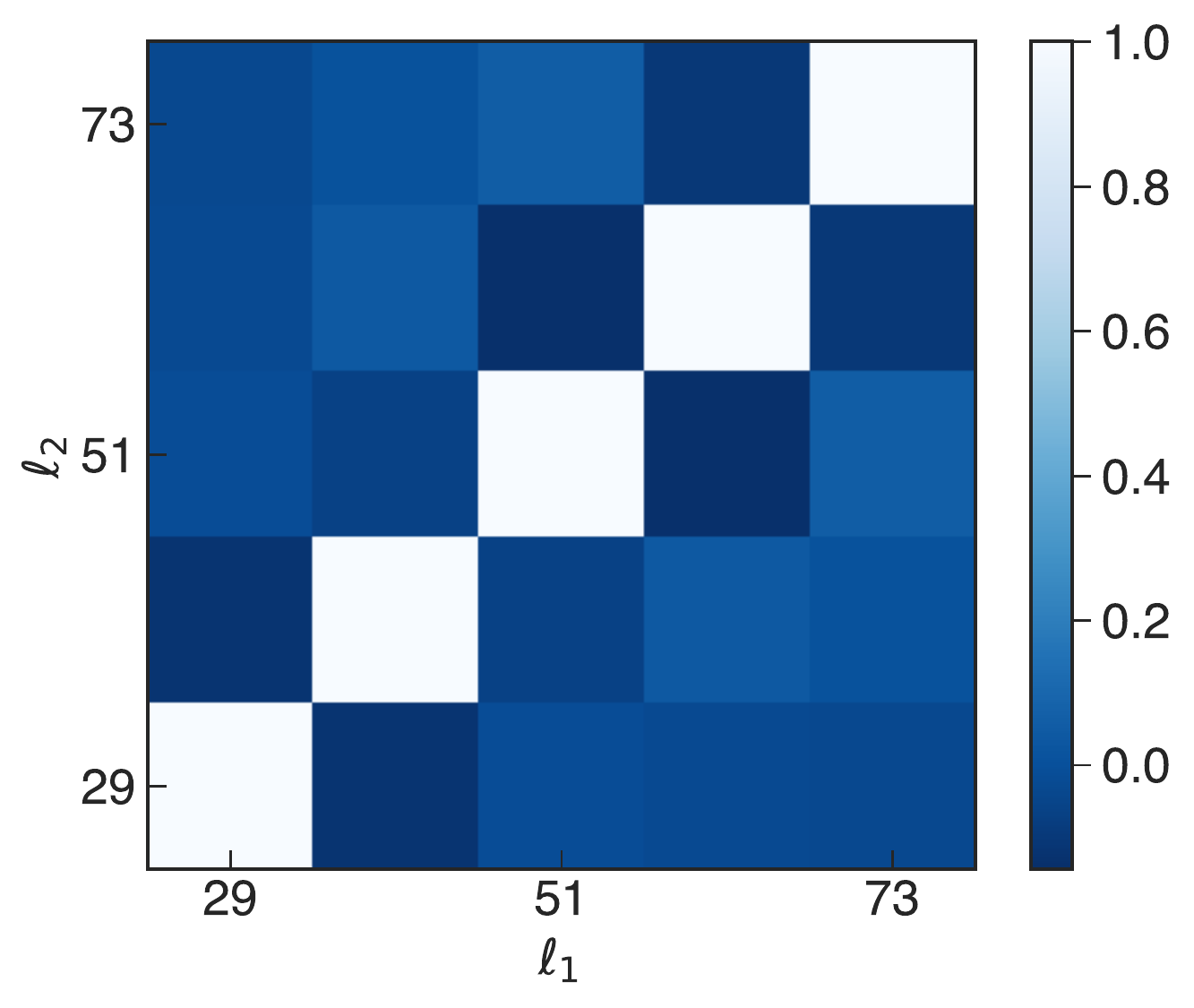}
\includegraphics[width=0.23\textwidth]{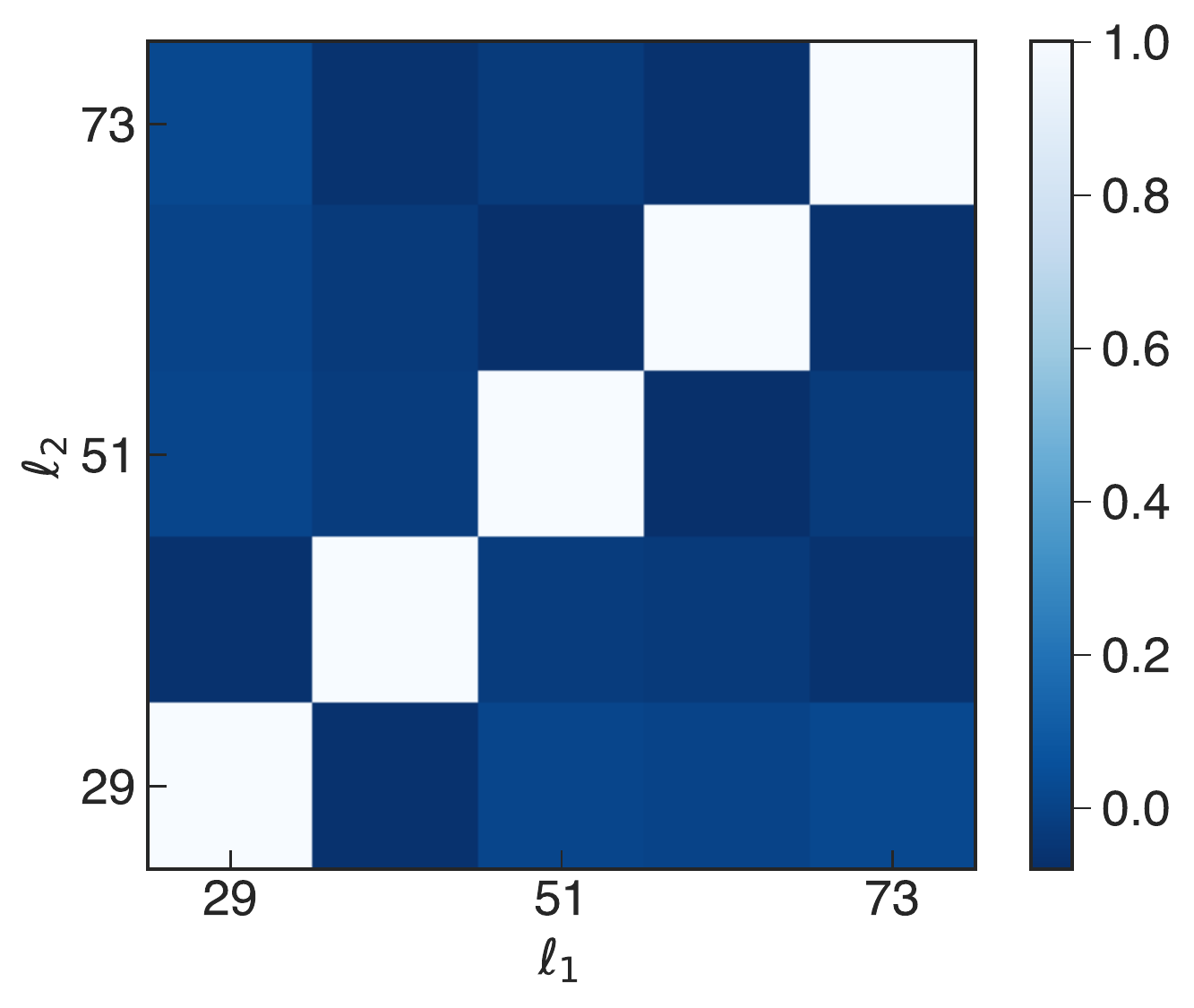}
\caption{Normalized covariance matrices $\mathcal{C}_{\ell, \ell'} = {\rm cov}(\hat{C}_\ell, \hat{C}_{\ell'}) /\sqrt{{\rm var}(\hat{C}_\ell){\rm var}(\hat{C}_{\ell'})}$ of the two QML methods for the space-based experiment with inhomogeneous noise. The matrices are obtained from estimates of 1000  simulations for classic QML estimator (upper panel) and for QML-SZ estimator (lower panel). The left diagrams show the covariance matrices of $TB$ mode, the right diagrams show equivalent plots of $EB$ mode. }
\label{fig:AliCPT_cov}
\end{figure}

In Fig.\ref{fig:AliCPT_cov}, we have shown the normalized covariance matrices for the power spectra estimates with two QML methods, the first two pictures are the covariance matrix of $TB$ and the last two pictures are the covariance matrix of $EB$. For the ground-based experiment, we also find our covariance matrices to be approximately diagonal, showing that the band power leakages have been suitably removed.

\subsection{Computational performance}
\label{sub:computation}

The computational complexity of the QML estimator is $O(N^3_d)$, where $N_d$ is the length of the data vector. \citet{chen2021fast} has shown that $N_d$ for the QML-SZ method is one third of the $N_d$ for the standard QML estimator. This significantly reduces the computational requirements for the QML-SZ method. In Tables \ref{tab:perf_tab_72} and \ref{tab:perf_tab_12}, we summarize the computational parameters for these two QML estimators for the space-based and ground-based experiment case, respectively. We run our computations on an Intel Xeon E2620 2.10 GHz workstation, and list the \texttt{NSIDE}, $N_d$, $\ell_{\rm max}$, RAM (in gigabytes), computation time for a single computation.

%
\begin{table}[h]
\caption{Performance comparison for different estimators in space-based experiment example. \label{tab:perf_tab_72}}
\centering
\begin{tabular}{l c c c c c}   
\tableline
 Estimator & \text{NSIDE} & $N_d$ & $\ell_{max}$ & RAM (GB) & Time ($s$)  \\ 

\tableline
QML 					& 16 & 6681 & 47 & 158.8  & $\sim$ 28527 \\ 
QML-SZ 					& 16 & 2139 & 47 & 3.6   & $\sim$ 98  \\ 
\tableline
\end{tabular}

\end{table}

As shown in Table \ref{tab:perf_tab_72}, for the QML-SZ estimators, the computation time is only about $1/280$ of that for classic QML estimator. This happens because the data vector size, $N_d$ for the scalar QML method is about $1/3$ of that for the classic QML method. Additionally the QML-SZ method are based on the scalar mode QML method, its algorithm complexity lower than the full form standard QML method, thereby the computation is faster.
%
\begin{table}[h]
\caption{Performance comparison for different estimators in ground-based experiment example. \label{tab:perf_tab_12}}
\centering
\begin{tabular}{l c c c c c}   
\tableline
 Estimator & \texttt{NSIDE} & $N_d$ & $\ell_{max}$ & RAM (GB) & Time ($s$)  \\ 

\tableline
QML 					& 32 & 5256 &  95 & 21.8  & $\sim$ 1320  \\ 
QML-SZ 					& 32 & 1752  &  95 & 1.58  & $\sim$ 28   \\ 
\tableline
\end{tabular}

\end{table}

In Table \ref{tab:perf_tab_12}, we show the same set of parameters for the ground-based experiment case. In this example, we compute QML methods at $\texttt{NSIDE}=32$. In Table \ref{tab:perf_tab_12}, we show that the  QML-SZ method saves on both computation time and memory requirements. Comparing with the results listed in the Table \ref{tab:perf_tab_72}, we find that with increase in $N_d$, the advantages of the QML-SZ method in computing time becomes obvious. 
For the ground-based case, the QML-SZ method is near-optimal in the entire multipole range of interest, and their computational requirements imply that they can be applied on higher resolution maps to compute the power spectrum at higher multipoles.


\section{Discussions and Conclusions}
\label{sec:conclusion}

In this work, we present one novel unbiased alternative estimator to estimate the cross-correlation power spectrum, called QML-SZ estimator. We using SZ method to solve the E-to-B leakage problem exist in CMB polarized fields and obtain scalar $E$-mode map, $\mathcal{E}(\hat{n})$, and $B$-mode map, $\mathcal{B}(\hat{n})$. By using the SZ method, we eliminate the coupling relationship between the polarization fields and decompose polarization $QU$ maps into independent pure $E$-mode map, $\mathcal{E}(\hat{n})$, and pure $B$-mode map, $\mathcal{B}(\hat{n})$. After that, all CMB information is contained in three scalar map, $T(\hat{n})$, $\mathcal{E}(\hat{n})$ and $\mathcal{B}(\hat{n})$. Through the scalar QML estimator we can reconstruct cross-correlation power spectrum from these scalar maps.  

We apply three different estimators to both space-based CMB experiment and ground-based CMB experiment, the details of simulation steps can be found in Sec. \ref{sec:SIMULATION_SETUP}. The final results of the $TE$ power spectrum showed in Appendix \ref{apdx:TE}. For the Planck case,  we find that the errors of the QML-SZ method are significantly larger than the hybrid pseudo-$C_\ell$ estimator and standard QML estimator for $\ell \le 5$, while it performs near-optimally for the rest of the multipole range. \textbf{As to the ground-based CMB experiment, the performance of hybrid pseudo-$C_\ell$ as well as the other two QML methods.} For the space-based CMB experiment scenario, all $TB$ and $EB$ power spectrum reconstructed by QML estimators has better performance on reducing the error bars than hybrid pseudo-$C_\ell$ estimator. For $\ell \le 5$ part, the results of QML-SZ still have significant large than standard QML methods.  The main reason is that the input power spectrum of QML-SZ is related to the factor $N_\ell$, which leads to the low multipoles being more sensitive to the impact of the high multipole. In order to improve the performance of the QML-SZ estimator on large scale, we need to present a better filter technology in our follow-up work. For ground-based CMB experiment, \textbf{the performance of hybrid pseudo-$C_\ell$ to estimate $TB$ and $EB$ power spectrum as well as the other two QML estimators too.} That means hybrid pseudo-$C_\ell$ is the best choice in most realistic ground-based CMB experiment scenarios.

Compared with the traditional QML estimator, the pixel number of the QML-SZ estimator is just $\sim 1/3$ of that of the traditional QML estimator, which greatly reduces the computational requirements (both the memory requirement and computation time). 

In conclusion, we present a new estimator for estimating the cross-correlation power spectrum and compare it with the other two classic methods hybrid pseudo-$C_\ell$ estimator and standard QML estimator in this article. We found that for the ground-based CMB experiment, the performance of pseudo-$C_\ell$ estimator a is good enough, which error bars are very close to the optimal errors and also have obvious advantages in calculation speed, we do not need to develop a QML-based estimator. The biggest advantage of our method is used to estimate $TB$ and $EB$ power spectrum in space-based cases. In this situation, the QML-SZ estimator has 
better performance than hybrid pseudo-$C_\ell$ estimator in the entire multipole range of analysis. While the errors of the QML-SZ results are not as small as the standard QML method for the lowest few multipoles, considering the QML-SZ estimator saves a lot of running time and memory, it is still worth us continued research to improve its performance. We can study new filtering algorithms or try more degraded schemes to suppress the influence of high multipoles in future work.

\acknowledgments{ We would like to thank AliCPT pipeline task force for helpful discussions. {Some of the results in this paper have been derived using the HEALPix \citep{2005ApJ...622..759G}}. This work is supported by the National Key R\&D Program of China Grant No. 2021YFC2203100, NSFC No. 11903030, the Fundamental Research Funds for the Central Universities under Grant No. WK2030000036 and WK3440000004, Key Research Program of the Chinese Academy of Sciences, Grant No. XDPB15, and the science research grants from the China Manned Space Project with NO.CMS-CSST-2021-B01 and NO. CMS-CSST-2021-B11.}

\appendix 
\section{$TE$ POWER SPECTRUM}
\label{apdx:TE}

\begin{figure*}[h]
\centering
\includegraphics[width=\textwidth]{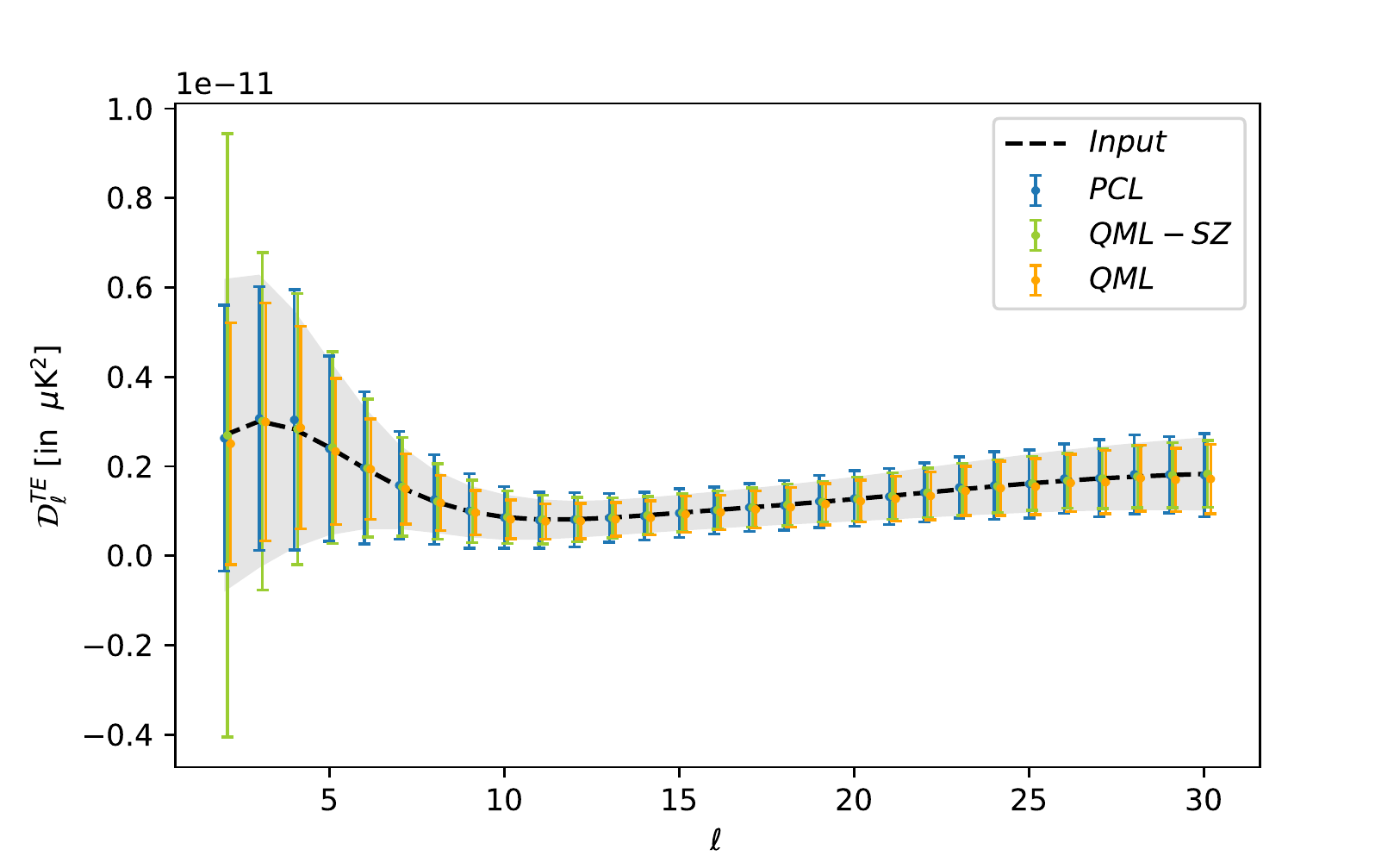}
\caption{Plot of the results for $TE$-mode power spectrum estimates for the realistic space-based CMB experiment. The observed sky is simulated at \texttt{NSIDE}=512 with $\ell_{\rm max}$=1024 and the noise levle is set to 3$\mu K$-arcmin. The input $TE$-mode power spectrum is shown with the black curve. The classic QML method results are shown with orange line, QML-SZ method results with green line. These results are computed at \texttt{NSIDE}=16. We also show PCL estimator results, obtained with NaMaster, with $\delta_c = 6^\circ$ C2 apodization, with blue line. The gray region denotes the analytical approximation of the error bounds. The data points are the mean of 1000 estimates, and the error bars are given by the standard deviation of the estimates.}
\label{fig:Planck_TE}
\end{figure*}

The $TE$ power spectrum does not involve the $E$-$B$ leakage problem, so the PCL estimator is adequate in most situations. For the space-based experiment case, we show the final results of three estimators in Fig. \ref{fig:Planck_TE}.

We find that all three methods give unbiased estimates for the $TE$ band powers. The standard QML method has nearly-optimal error bars throughout the entire multipole range, PCL method has sub-optimal error bars. For the QML-SZ estimator, the error bars for $\ell \le 5$ show a significant increase. The power spectrum of scalar $E$-map $\mathcal{E}(\hat{n})$ is $C_\ell^{\mathcal{EE}}=N^2_{\ell}C^{EE}_\ell$, due to the existence of coefficient factors $N^2_\ell$, the power from high multipoles leaks to the lower multipoles and increases the uncertainty on the large angular scales. Based on the current framework, in order to solve this problem, we need to present a better downgrade method and we will try to do this in our follow-up work. Here we just show a new possible method of estimating the $TE$ power spectrum.

\begin{figure}[h]
\centering
\includegraphics[width=0.8\textwidth]{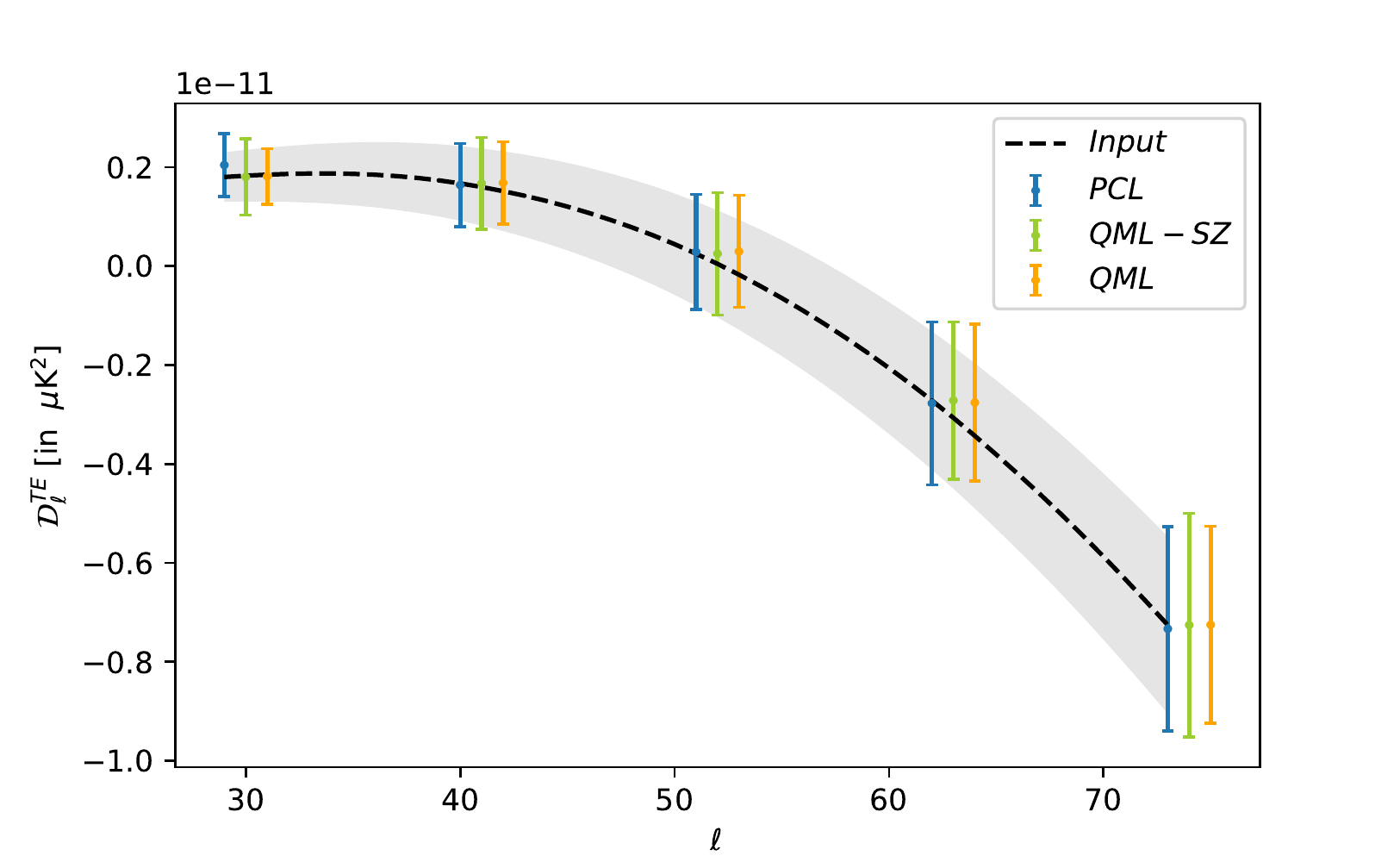}
\caption{Plot of the results for $TE$-mode power spectrum estimates for the realistic ground-based CMB experiment. The observed sky is simulated at \texttt{NSIDE}=512 with $\ell_{\rm max}$=1024 and the noise level is set to 3$\mu K$-arcmin. The input $TE$-mode power spectrum is shown with the black curve. The classic QML method results are shown with orange line, and QML-SZ method results are shown with green curve. These results are computed at \texttt{NSIDE}=32. We also show PCL estimator results, obtained with NaMaster, with $\delta_c = 10^\circ$ C2 apodization, with blue line. The gray region denotes the analytical approximation of the error bounds. The data points are the mean of 1000 estimates, and the error bars are given by the standard deviation of the estimates.}
\label{fig:AliCPT_TE}
\end{figure}

The results of the ground-based experiment case are shown in Fig. \ref{fig:AliCPT_TE}. \textbf{We find that all three methods give the quite similar results. All the estimators are unbiased, and the error bars are close to each other for every multipole.} As well known, the PCL method is much faster than the QML-based methods, so we do not need to develop a new calculation estimator for the small-scale case to reconstruct $TE$ band powers.


\begin{figure}[h]
\centering
\includegraphics[width=0.44\textwidth]{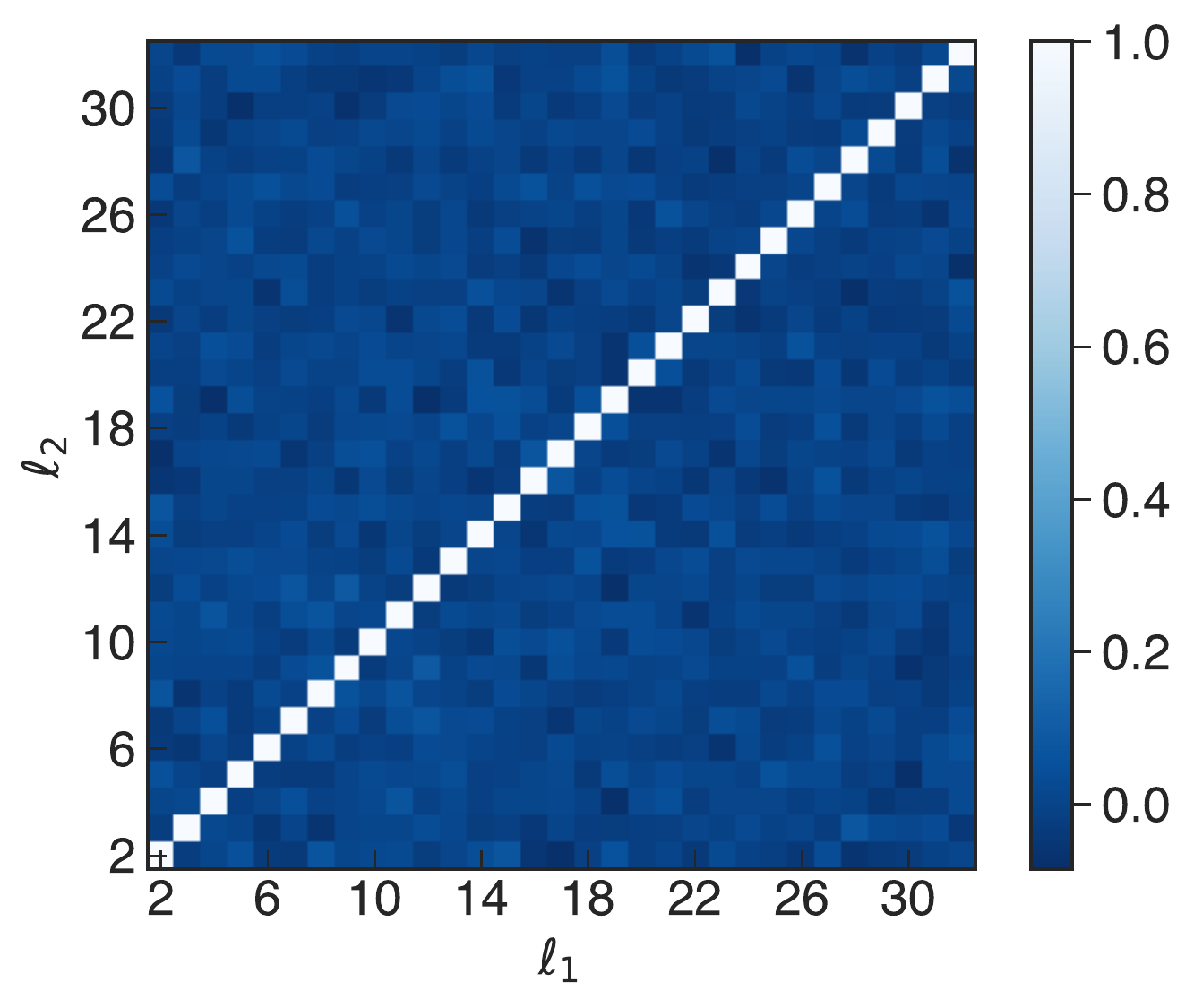}
\includegraphics[width=0.44\textwidth]{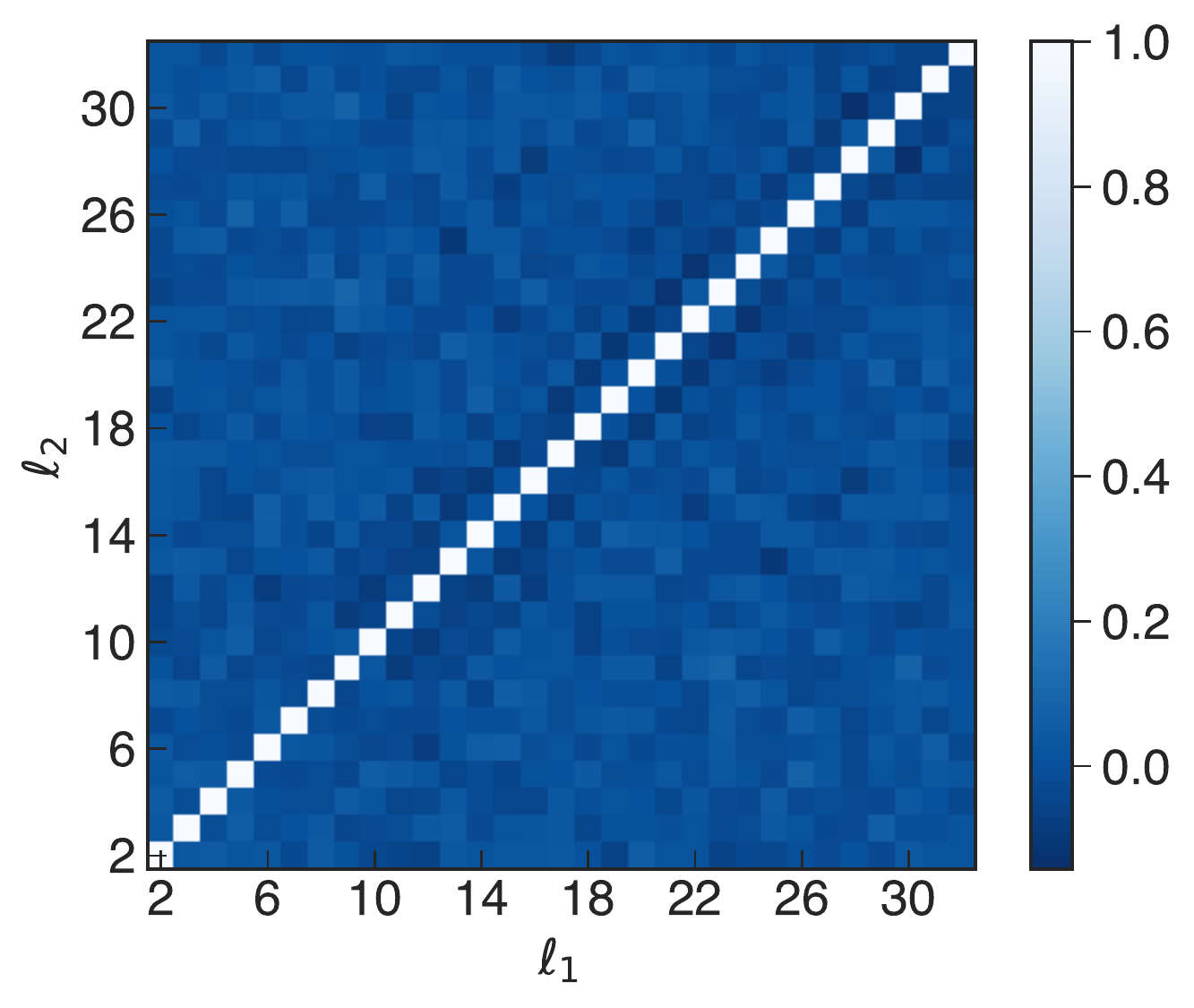}
\caption{Normalized covariance matrices $\mathcal{C}_{\ell, \ell'} = {\rm cov}(\hat{C}_\ell, \hat{C}_{\ell'}) /\sqrt{{\rm var}(\hat{C}_\ell){\rm var}(\hat{C}_{\ell'})}$ of the QML methods for the space-based experiment. The matrices are obtained from estimates of 1000  simulations for classic QML estimator (left) and for QML-SZ estimator (right). }
\label{fig:cov_TE}
\end{figure}
\begin{figure}[ht]
\centering
\includegraphics[width=0.44\textwidth]{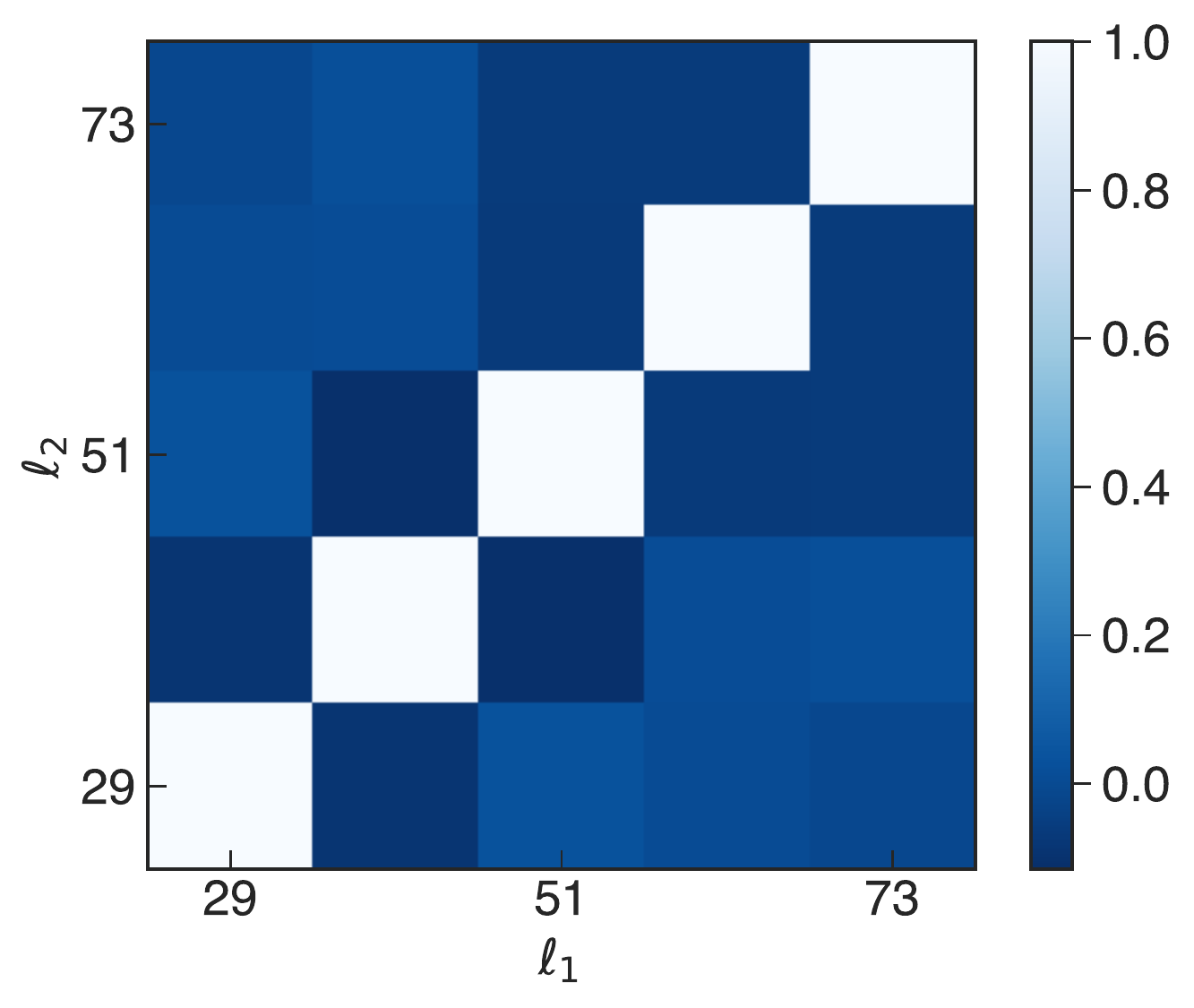}
\includegraphics[width=0.44\textwidth]{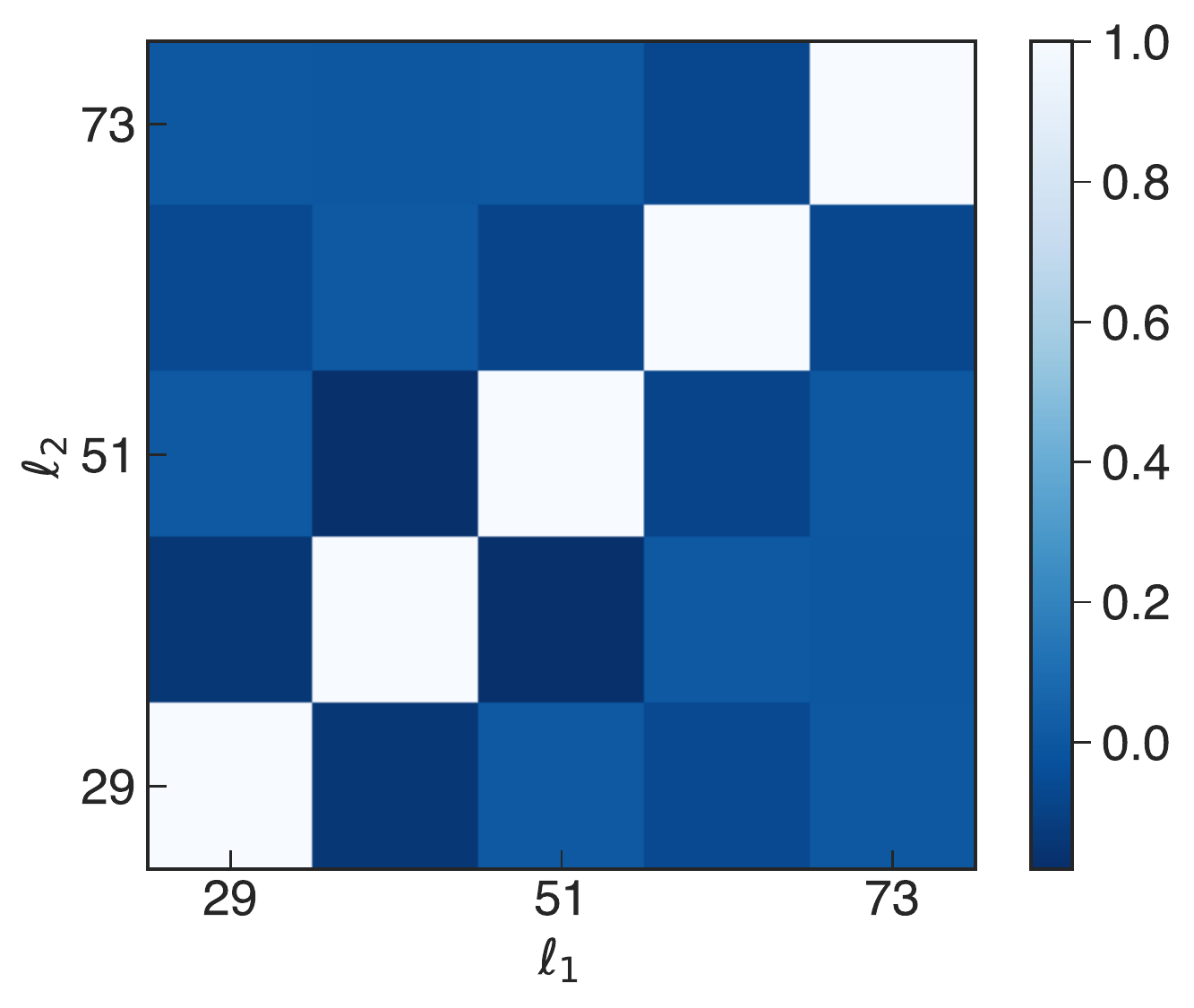}
\caption{Normalized covariance matrices $\mathcal{C}_{\ell, \ell'} = {\rm cov}(\hat{C}_\ell, \hat{C}_{\ell'}) /\sqrt{{\rm var}(\hat{C}_\ell){\rm var}(\hat{C}_{\ell'})}$ of the QML methods for the ground-based experiment. The matrices are obtained from estimates of 1000  simulations for classic QML estimator (left) and for QML-SZ estimator (right). }
\label{fig:AliCPT_cov_TE}
\end{figure}
Finally, the normalized covariance matrices for the power spectra estimates with the QML methods shown in Fig. \ref{fig:cov_TE} and in Fig. \ref{fig:AliCPT_cov_TE} and as you can see that the band power leakages have been suitably removed. 

\bibliographystyle{aasjournal}
\bibliography{biblio}

\end{document}